\input epsf
\documentstyle[preprint,eqsecnum,aps]{revtex}

\def\INSERTCAP#1#2{\vbox{%
{\narrower\noindent%
\multiply\baselineskip by 3%
\divide\baselineskip by 4%
{\rm Table #1. }{\sl #2 \medskip}}
}}

\def\btod{ $\bar B \to D \ell \bar \nu$}
\def\btods{ $\bar B \to D^* \ell \bar \nu$}

\begin{document}

\preprint{\vbox{
\hbox{CMU-HEP 97--07}\hbox{UCSD/PTH 97--12}
}}

\title{Precision Corrections to Dispersive Bounds on Form Factors}

\author{C. Glenn Boyd\footnote{boyd@fermi.phys.cmu.edu}}

\address{Department of Physics, Carnegie Mellon University}

\author{Benjam\'{\i}n Grinstein\footnote{bgrinstein@ucsd.edu} \\
{\it and} \\ Richard F. Lebed\footnote{rlebed@ucsd.edu}}

\address{Department of Physics,
University of California at San Diego, La Jolla, CA 92093}
\date{May, 1997}
\maketitle
\begin{abstract}

	We present precision corrections to dispersion relation bounds
on form factors in bottom hadron semileptonic decays and analyze their
effects on parameterizations derived from these bounds.  We
incorporate QCD two-loop and nonperturbative corrections to the
two-point correlator, consider form factors whose contribution to
decay rates is suppressed by lepton mass, and implement more realistic
estimates of truncation errors associated with the parameterizations.
We include higher resonances in the hadronic sum that, together with
heavy quark symmetry relations near zero recoil, further tighten the
sum rule bounds.  Utilizing all these improvements, we show that each
of the six form factors in \btod\ and \btods\ can be described with
$3\%$ or smaller precision using only the overall normalization and
one unknown parameter.  A similar one-coefficient parameterization of
one of the $\Lambda_b \rightarrow \Lambda_c \ell \bar \nu$ form
factors, together with heavy quark symmetry relations valid to order
$1/m^2$, describes the differential baryon decay rate in terms of one
unknown parameter and the phenomenologically interesting quantity
$\bar \Lambda_\Lambda \approx M_{\Lambda_b} - m_b$.  We discuss the
validity of slope-curvature relations derived by Caprini and Neubert,
and present weaker, corrected relations.  Finally, we present sample
fits of current experimental \btods\ and \btod\ data to the improved
one-parameter expansion.
\end{abstract}



\narrowtext

\section{Introduction} \label{int}

	The decays of bottom hadrons provide fertile ground for
explorations of both weak and strong interactions.  The empirical
smallness of the CKM elements $|V_{cb}|$ and $|V_{ub}|$ implies that
$b$-hadron decays are relatively long-lived, and thus more easily
analyzed, while the heaviness of the $b$ quark means that the heavy
quark effective theory (HQET)\cite{IW,HQET} provides a reliable
expansion for describing the approximate decoupling of the $b$ quark
from the rest of the hadron, leading to a substantial simplification
of the strong-interaction dynamics.

	Semileptonic decays of $b$-hadrons are particularly tractable
from the theoretical point of view, since the leptonic current may be
trivially separated from the hadronic current.  Furthermore, focusing
upon exclusive decays permits one to avoid questions regarding the
validity of quark-hadron duality near kinematic endpoint regions.  On
the other hand, exclusive processes must be described in terms of a
number of nonperturbative form factors that encapsulate the physics of
the hadronization process.

	HQET has provided a substantial leap forward in the exclusive
analysis, demonstrating that heavy-to-heavy quark transitions at zero
recoil are accompanied by a complete overlap of initial- and
final-state hadron wave functions\cite{IW,Luke}.  Consequently, each
form factor possesses a well-defined normalization at this kinematic
point.  For the case of $\bar B^{(*)} \rightarrow D^{(*)}$
transitions, for example, each of the 20 form factors either vanishes
or is proportional to a single universal form factor $\xi$, which
represents this wave function overlap as a function of momentum
transfer.  Sum rule constraints restrict the slope and magnitude of
these form factors\cite{BJ,Volo,BR}, but otherwise shapes of the form
factors are {\it \`{a} priori\/} undetermined functions of momentum
transfer.

	However, some recent
work\cite{BL,BGLNP,BGLPL,BGLPRL,BS,Lel,Bec1,Bec2,Cap,CN,CM,CM2} has
demonstrated that one may obtain rather nontrivial and
model-independent constraints on the shape of such form factors in the
context of dispersion relations.  Using basic field theoretic
properties like causality and crossing symmetry, one finds that the
shape of a given form factor is determined to high accuracy over its
entire kinematic range by its value at only a small number of
points\cite{BL,BGLNP,BGLPL,BGLPRL}.  Indeed, for the case of $\bar B
\rightarrow D^{(*)} \ell \bar \nu$, the HQET normalization plus only
one additional point determines most of the form factors to 3--7\%
accuracy\cite{BL}.  Similar analyses provide interesting and useful
constraints on $\bar B \rightarrow \pi \ell \bar
\nu$\cite{BL,BGLPRL,BS,Lel,Bec1}, $\bar B \rightarrow \rho \ell
\bar \nu$\cite{Bec2}, and $\Lambda_b \rightarrow \Lambda_c \ell \bar
\nu$\cite{BL}.

	At the core of this analysis lies a dispersion relation in
momentum space that relates the integral of an inclusive production
rate to a two-point function evaluated far from physical resonances,
where perturbative QCD is quantitatively reliable.  The production
rate is a sum of positive-definite exclusive rates; it follows that
all contributions to the hadronic side serve to saturate more and more
closely the bound from the perturbative side.  Clearly, two
improvements to this procedure include a better calculation of the QCD
side (which includes both multi-loop effects and nonperturbative
vacuum condensates), and the inclusion of more states on the hadronic
side. The former may be extracted from existing literature, while the
latter may be achieved using heavy quark symmetry relations.  The
inclusion of higher states has previously been investigated for
elastic $B \to B$ scattering\cite{CM}\ and for $\bar B \to \pi \ell
\bar \nu$ decays\cite{BS}, but not for $b \to c$ transitions.
For mesonic and baryonic $b \to c$ transitions, the inclusion of
higher states improves considerably the precision of the
dispersion-relation constraints, and represents the primary
contribution of this work.  In addition, we include in this analysis
nonzero lepton masses, anticipating future measurements of processes
such as $\bar B \rightarrow D^{(*)} \tau \bar \nu_\tau$ or $\bar B
\rightarrow \pi \tau \bar \nu_\tau$.

	This paper is organized as follows.  In Sec.\ \ref{ff} we
define the form factors for the physically-observed semileptonic
decays and present expressions for differential decay widths including
lepton masses.  Section \ref{rev} presents a brief review of the
dispersive method for obtaining bounds on the form factors.
Ingredients of the form factor parameterizations, including explicit
formulas for the QCD results at one and two loops as well as leading
nonperturbative effects, appear in Sec.\ \ref{param}, followed by a
tabulation of the form factor weighting functions, which are central
to the dispersive method.  In Sec.\ \ref{add} we discuss the inclusion
of additional, previously ignored hadronic states into the dispersion
relations, and show how heavy quark effective theory may be used to
include their effects and thereby tighten form factor constraints.
Section \ref{errs} defines and presents the truncation errors that
measure the quality of our form factor parameterizations. In Sec.\
\ref{CNsec} we discuss slope-curvature relations derived by Caprini
and Neubert\cite{CN}, point out an invalidating assumption, and
examine the form of the corrected relations.  In Sec.\ \ref{fit} we
present the results of the current analysis in fits to the latest
experimental data, and in Sec.\
\ref{conc} we conclude.

\section{Form Factors} \label{ff}

	We begin by defining form factors for the semileptonic decays
of interest in terms of hadronic transition matrix elements.  From the
field theory point of view, it is most convenient to define form
factors as coefficients of independent Lorentz structures appearing in
the matrix element. However, the combinations of form factors most
easily obtained from data are those appearing in a sum of squares in
the differential rates, namely, the helicity amplitudes.  As seen
below, the helicity amplitudes are particular linear combinations of
the original form factors, and thus simply form a different basis for
the description of the matrix elements. It is these helicity amplitudes
we wish to constrain.

	The notation used throughout the paper is as follows: The
generic $b \rightarrow c$ semileptonic decay is denoted by $H_b \to
H_c \ell \bar \nu$, where the hadron $H_b$ has mass $M$ and momentum
$p$, the daughter hadron $H_c$ has mass $m$ and momentum $p'$, and the
charged lepton $\ell$ has mass $m_\ell$. The momentum transfer $t =
(p-p')^2$ is the invariant mass-squared of the lepton pair (or virtual
$W$).  The polar vector and axial vector flavor-changing currents are
denoted by $V^\mu = \bar c \gamma^\mu b$ and $A^\mu = \bar c
\gamma^\mu \gamma_5 b$.  Finally, it is convenient to define the
kinematic invariants
\begin{eqnarray} \label{kdef}
t_\pm & = & (M \pm m)^2  , \nonumber \\
k & = & M \sqrt{{\bf p}^2 \over t} \equiv  
 \sqrt{  (t_+ -t)(t_--t) \over 4 t } ,
\end{eqnarray}
where ${\bf p}$ is the three-momentum of $H_c$ in the rest frame
of $H_b$.  Note that the identity of the decaying quark does not enter
into the expressions below except through hadron masses, so the
same expressions apply to such decays as $\bar B \rightarrow \pi \ell
\bar \nu$, $D \rightarrow \overline{K}^{\, *} \ell \nu$, and so on;
only the spins of the hadrons are relevant.

	The inclusion of charged lepton masses brings into the
differential widths a new set of helicity amplitudes, and thus a new
laboratory for studying the strong interaction.  The contributions of
such terms is suppressed by a factor of $m_\ell^2$, and arises through
a virtual $W$ with the quantum numbers of a scalar, {\it i.e.,} a
timelike polarization. Angular momentum conservation forbids the decay
of such a state to a right-handed antineutrino and a left-handed {\it
massless\/} charged lepton, so such decays are necessarily accompanied
by a helicity-suppression factor $m_\ell^2$.  Although the detection
of lepton mass-dependent effects is presently beyond the means of
current experiments, the gradual accumulation of statistics may
eventually make such effects discernible in decays like $\bar B
\rightarrow D^{(*)} \tau \bar \nu$ or $\bar B \rightarrow \pi \tau
\bar \nu$.

	Finally, it should be noted that the differential widths
$d\Gamma/dt$ presented below have already been integrated in lepton
energy, or equivalently over angles of final-state particles.  This
reflects only the current thrust of experiment and does not indicate a
limitation of the dispersive method described below.  Indeed, were the
statistics available, it would be interesting to consider the double
differential decay distribution $d^2\Gamma/dE_\ell dt$, for then one
could probe the parity-violating interference terms between vector and
axial vector weak currents.  In terms of the QCD side of the
dispersion relations described below, one would also need to compute
$V$-$A$ correlators, in addition to $V$-$V$ and $A$-$A$.

\subsection{\boldmath $\bar B \rightarrow D \ell \bar \nu$}

	The hadronic matrix element governing the rate of $\bar B
\rightarrow D \ell \bar \nu$ may be described by form factors
\begin{equation}
\langle D(p') | V^\mu | \bar B(p) \rangle = f_+ (p+p')^\mu +
f_- (p-p')^\mu \ ,
\end{equation}
that enter the differential rate as
\begin{equation}
\frac{d\Gamma}{dt} = \frac{G_F^2 |V_{cb}|^2}{192\pi^3 M^3}
\frac{k}{t^{\frac 5 2}} \left( t-m_\ell^2 \right)^2
\Biggl[ 4 k^2 t \left( 2t+m_\ell^2 \right) |f_+|^2 + 3m_\ell^2
|f_0|^2 \Biggr] ,
\end{equation}
where
\begin{eqnarray}
f_0 (t) & \equiv & (M^2-m^2) f_+ (t) + t f_- (t).
\end{eqnarray}
\subsection{\boldmath $\bar B \rightarrow D^* \ell \bar \nu$}

	The matrix elements for $\bar B \rightarrow D^* \ell \bar \nu$
depend on four form factors,
\begin{eqnarray}
\langle D^* (p',\epsilon) | V^\mu | \bar B(p) \rangle & = & ig
\epsilon^{\mu \alpha \beta \gamma} \epsilon^*_\alpha p'_\beta p_\gamma
, \nonumber \\
\langle D^*(p',\epsilon) | A^\mu | \bar B(p) \rangle & = & f
\epsilon^{* \mu} + (\epsilon^* \cdot p) \left[ a_+ (p+p')^\mu + a_-
(p-p')^\mu \right] ,
\end{eqnarray}
that enter the differential rate in the combinations 
\begin{eqnarray}
\frac{d\Gamma}{dt} & = & \frac{G_F^2 |V_{cb}|^2}{192\pi^3 M^3}
\frac{k}{t^{\frac 5 2}} \left( t - m_\ell^2 \right)^2
\nonumber \\ & & \times \Biggl\{ \left( 2t + m_\ell^2 \right)
\left[ 2t |f|^2 + |{\cal F}_1|^2+ 2 k^2 t^2 |g|^2 \right] +
3m_\ell^2 k^2 t |{\cal F}_2|^2 \Biggr\} ,
\end{eqnarray}
where
\begin{eqnarray} \label{F12}
{\cal F}_1 (t) & \equiv & \frac 1 m \Biggl[ 2k^2 t a_+ (t) - \frac 1
2 (t-M^2+m^2) f (t) \Biggr] , \nonumber \\ {\cal F}_2 (t) & \equiv &
\frac 1 m \Biggl[ f (t) + (M^2-m^2) a_+ (t) + t a_- (t)
\Biggr] .
\end{eqnarray}
\subsection{\boldmath $\Lambda_b \rightarrow \Lambda_c \ell \bar \nu$}

	This decay can be described by six form factors, defined by
\begin{eqnarray}
\langle \Lambda_c (p') | V^\mu | \Lambda_b (p) \rangle & = & \bar u_c
(p') \left[ F_1 \gamma^\mu + F_2 v^\mu + F_3 v'^\mu \right] u_b(p),
\nonumber \\ \langle \Lambda_c (p') | A^\mu | \Lambda_b (p) \rangle &
= & \bar u_c (p') \left[ G_1 \gamma^\mu + G_2 v^\mu + G_3 v'^\mu
\right] \gamma_5 u_b(p),
\end{eqnarray}
with $v = p/M_{\Lambda_b}$ and $v' = p'/M_{\Lambda_c}$.  This gives
\begin{eqnarray}
\frac{d\Gamma}{dt} & = & \frac{G_F^2 |V_{cb}|^2}{192\pi^3 M^3}
\frac{k}{t^{\frac 5 2}} \left( t-m_\ell^2 \right)^2 \nonumber
\\ & & \times \Biggl\{ \left( t_- -t  \right) \left( 2t + m_\ell^2 \right)
\left[ 2t |F_1|^2 + |H_V|^2 \right] + 3m_\ell^2
\left( t_+ -t \right) |F_0|^2 \Biggr. \nonumber \\ & & + \hspace{1.5ex}
\Biggl. \left( t_+ -t \right) \left( 2t + m_\ell^2 \right)
\left[ 2t |G_1|^2 + |H_A|^2 \right] + 3m_\ell^2
\left( t_- -t \right) |G_0|^2 \Biggr\} ,
\end{eqnarray}
where
\begin{eqnarray}
H_V (t) & = & (M+m)F_1 + \frac{1}{2}(t_+ -t) \left(
\frac{F_2}{M} + \frac{F_3}{m} \right) , \nonumber \\
H_A (t) & = & (M-m)G_1 - \frac{1}{2}(t_- -t) \left( \frac{G_2}{M}
+ \frac{G_3}{m} \right) , \nonumber \\
F_0 (t) & = & (M-m)F_1 + \frac{1}{2M}
\left( t + M^2 - m^2 \right) F_2 - \frac{1}{2m}
\left( t - M^2 + m^2 \right) F_3 , \nonumber \\
G_0 (t) & = & (M+m)G_1 - \frac{1}{2M}
\left( t + M^2 - m^2 \right) G_2 + \frac{1}{2m}
\left( t - M^2 + m^2 \right) G_3 .
\end{eqnarray}

\section{Review of the Dispersive Approach} \label{rev}

	Constraints on a generic $H_b \to H_c \ell \bar \nu$ form
factor $F(t)$ are obtained by noting that the amplitude for production
of $H_b \bar H_c$ from a virtual $W$ boson is determined by the
analytic continuation of $F(t)$ from the semileptonic region of
momentum-transfer $m_\ell^2 \le t \le t_- $ to the pair-production
region $t_+ \le t$.  The idea of the dispersion relation is to
constrain $F(t)$ in the pair-production region using perturbative QCD,
then use analyticity to translate that constraint into one valid in
the semileptonic region\cite{BMdeR}.  A detailed derivation can be
found in
\cite{BL,BGLNP,BS}; here we merely outline the essential elements.

	In QCD, the two-point function\footnote{This definition
differs slightly from that used in \cite{BL,BGLNP,BGLPL,BGLPRL,BS},
and serves to separate $\Pi^{\mu\nu}$ into mani\-festly spin-0 and
spin-1 pieces.  Then the functions $\chi^T$, $\chi^L$ defined in
(\ref{chis}) coincide with $\chi$, $\chi^L$ defined in the previous
works.} of a flavor-changing current $J = V,A,$ or $V-A$
\begin{equation} \label{pol}
\Pi^{\mu \nu}_J (q) = \frac{1}{q^2}(q^\mu q^\nu-q^2 g^{\mu\nu})
\Pi_J^T(q^2) + \frac{q^\mu q^\nu}{q^2} \Pi_J^L(q^2)
\equiv i \int d^4 \! x \, e^{iqx} \langle 0 |
{\rm T} J^\mu(x) J^{\dagger\nu}(0) |0 \rangle ,
\end{equation}
is rendered finite by making one or two subtractions, leading to the
dispersion relations
\begin{eqnarray} \label{chis}
\chi^L_J (q^2) \equiv \frac{\partial\Pi_J^L}{\partial q^2} & = &
{1\over\pi}\int_0^\infty dt \, \frac{{\rm Im}\,\Pi_J^L (t)}{(t-q^2)^2}
, \nonumber \\
\chi^T_J (q^2) \equiv \frac{1}{2} \frac{\partial^2 \Pi^T_J}
{\partial (q^2)^2} & = & {1\over\pi}\int_0^\infty dt \, \frac{{\rm
Im}\,\Pi_J^T (t)}{(t-q^2)^3} .
\end{eqnarray}
The functions $\chi(q^2)$ may be computed reliably in perturbative QCD
for values of $q^2$ far from the kinematic region where the current
$J$ can create resonances: specifically,\hfil\\ $(m_b+m_c)\Lambda_{\rm
QCD}\ll (m_b+m_c)^2 -q^2$.  For $b \rightarrow c$, or $u$, $q^2=0$
satisfies this condition.

	Inserting a complete set of states $X$ into the two-point
function relates the $\Pi_J$ to the production rate of hadrons from a
virtual $W$,
\begin{equation} \label{ins}
{\rm Im}\,\Pi^{T,L}_J = \frac12 \sum_X (2 \pi)^4 
    \delta^4(q-p_X) | \langle 0| J | X \rangle |^2 \ ,
\end{equation}
where the sum is over all hadronic states $X$ with the same quantum
numbers as the current $J$, weighted by phase space.  Then, from the
dispersion relations (\ref{chis}), the perturbatively evaluated
$\chi(q^2)$ is equal to the integrated production rate of $W^* \to X$,
weighted with a smooth function of momentum transfer squared, $t$.
Since the sum is semipositive definite, one may restrict attention to
a subset of hadronic states to obtain a strict inequality.  In the
case of interest, we focus on $X$ being two-particle states of the
form $H_b \bar H_c$.  This places an upper bound on the form factor
$F(t)$ in the pair-production region that takes the form
\begin{equation}\label{tbd}
{1\over \pi \chi^T(q^2)} \int_{t_+}^\infty dt \, {W(t)\ |F(t)|^2 \over 
(t-q^2)^3 } \leq 1 ,
\end{equation}
from the Im $\Pi^T$ dispersion relations in Eq.~(\ref{chis}).  Here
$W(t)$ is a computable function of $t$ that depends on the particular
form factor under consideration.  A similar result holds for $\Pi^L$.

	Using analyticity to turn (\ref{tbd}) into a constraint in the
semileptonic region requires that the integrand is analytic below the
pair-production threshold $t < t_+$.  To do this, we introduce a
function
\begin{equation}\label{zblaschke}
z(t; t_s) =  { t_s - t \over (\sqrt{t_+ -t} + \sqrt{t_+ - t_s} )^2}
\end{equation}
that is real for $t_s < t_+$, zero at $t=t_s$, and a pure phase for $t
\ge t_+$. All the poles in the integrand of Eq.~(\ref{tbd}) can be
removed by multiplying by various powers of $z(t;t_s)$, provided the
positions $t_s$ of the sub-threshold poles in $F(t)$ are known.  Each
pole has a distinct value of $t_s$, and the product $z(t;t_{s1})
z(t;t_{s2}) \cdots$ serves to remove all of them.  Such poles arise as
the contribution of $B_c$ resonances to the form factor $F(t)$, as
well as singularities in the kinematic part of the integrand.  After
determining these positions phenomenologically, the upper bound on
$F(t)$ becomes
\begin{equation} \label{tdisp}
\frac{1}{\pi} \int_{t_+}^\infty dt \left| \frac{dz(t;t_0)}{dt} \right|
\cdot | \phi(t;t_0) P(t) F(t) |^2 \le 1 \ ,
\end{equation}
where the weight function $\phi(t; t_0)$ (known as an {\it outer
function} in complex analysis) is given by
\begin{equation}
 \phi(t;t_0) = \tilde P(t) \Biggl[ {W(t) \over 
  |dz(t;t_0)/dt| \, \chi^T(q^2) (t-q^2)^3 } \Biggr]^\frac12 \ .
\end{equation}
The factor $\tilde P(t)$ is a product of $z(t; t_s)$'s and $\sqrt{z(t;
t_s)}$'s, with $t_s$ chosen to remove the sub-threshold singularities
and cuts in the kinematic part of the integrand, while the {\it
Blaschke factor\/} $P(t)$ is a product of $z(t; t_p)$'s with $t_p$
chosen to be the positions of sub-threshold poles in $F(t)$.  The
functions $\phi(t; t_0)$ and $\tilde P (t)$ also depend on $q^2$,
which we leave implicit for notational simplicity, while $t_0$ is a
free parameter to be discussed in Sec.~\ref{param}.

	The quantity $\phi(t;t_0) P(t) F(t)$ may be expanded in a set
of orthonormal functions that are proportional to powers of $z(t;
t_0)$ [see (\ref{dispz}) below].  The function $z(t; t_0)$ has a
physical interpretation as a natural scale for the variation of $F(t)$
in the semileptonic region\cite{BS}, and will play an important role
throughout this paper.  We exhibit all relevant expressions in terms
of both variables, $z$ and $t$, in Sec.\ \ref{param}. The result of
the expansion in $z(t; t_0)$ is an expression for $F(t)$ valid even in
the semileptonic region,
\begin{equation} \label{masterparam}
F(t) = {1 \over P(t) \phi(t;t_0) } \sum_{n=0}^\infty
                   a_n \, z(t;t_0)^n \ \ ,
\end{equation}
where, as a result of Eq.~(\ref{tdisp}), the coefficients $a_n$ are
unknown constants obeying
\begin{equation} \label{abound}
 \sum_{n=0}^\infty  a_n^2 \le 1 \ .
\end{equation}
For the $b \to c$ processes that are the main subject of this paper,
$z(t;t_0)$ is no larger than $0.07$ for any physical momentum transfer
$m_\ell^2 \le t \le t_-$, and can be made substantially smaller by a
judicious choice of $t_0$, so the expansion can be truncated after the
first two or three terms.

\section{Parameterization Ingredients} \label{param}

	Generating a parameterization like (\ref{masterparam}) for a
particular form factor requires three ingredients: One needs the
perturbative evaluation of $\chi$ derived from the two-point function
for a current $J$, including the Wilson coefficients of
phenomenologically determined condensates.  In addition, the
functional form of the weighting function $\phi$ must be
computed. Finally, the masses of sub-threshold resonances with the
same quantum numbers as $J$ must be extracted from experiment or
potential models. The function $\phi$ depends on the form factor under
consideration, while $\chi$ and $P$ depend only on the current $J$.

\subsection{QCD Evaluation of \boldmath $\chi$}

	In the previous section we observed that it suffices to take
$q^2=0$ in the computation of $\chi_J^{T,L}$ for currents containing a
heavy quark.  This is convenient since then the perturbative
expressions become particularly simple.  Corrections to the
perturbative result may be included by expressing the two-point
function as an operator product expansion (OPE) and including the
leading nonperturbative vacuum condensates such as $\langle G^2
\rangle$ and $\langle \bar q q \rangle$; the total $\chi$ is the sum
of the perturbative and condensate terms,
\begin{equation}
 \chi = \chi_{\rm pert} + \chi_{\rm cond}. 
\end{equation}

	The full perturbative expressions to two loops may be obtained
through a lengthy but straightforward\footnote{We have corrected a
number of typographical errors in the perturbative results of
Ref.~\cite{GenP}\ and the Wilson Coefficient results of
\cite{RRY}\ to ensure compliance with various consistency
conditions. We also made several comparisons between references to
verify their agreement, once these corrections are made.}
manipulation of results existing in the literature\cite{GenP,DG}. At
$q^2=0$, $\chi_{\rm pert}$ is only a function of the ratio of quark
pole masses $u = m_c/m_b$, and for a vector current $J=V$ is given by
\begin{eqnarray} \label{Tpert}
\lefteqn{m_b^2 \chi^T_{{\rm pert}} (u) =}
\nonumber \\ & & \frac{1}{32 \pi^2 (1-u^2)^5}
\Bigl[ (1-u^2) (3+4u-21u^2+40u^3-21u^4+4u^5+3u^6) \nonumber \Bigr. \\ 
& & \Bigl. \hspace{6em} + 12u^3 (2-3u+2u^2)
\ln (u^2) \Bigr] \nonumber \\
& & + \frac{\alpha_s}{576\pi^3 (1-u^2)^6} \nonumber \\ & &
\hspace{2em}
\Bigl[ (1-u^2)^2
(75+360u-1031u^2+1776u^3-1031u^4+360u^5+75u^6) \nonumber \Bigr. \\ & &
\hspace{2em} +4u(1-u^2) (18-99u+732u^2-1010u^3+732u^4-99u^5+18u^6)
\ln (u^2) \nonumber \\ & & \hspace{2em}
+4u^3 (108-324u+648u^2-456u^3+132u^4+59u^5-12u^6-9u^7)
\ln^2 (u^2) \nonumber \\ & &
\hspace{2em} \Bigl. +8(1-u^2)^3 (9+12u-32u^2+12u^3+9u^4)
{\rm Li}_2 (1-u^2) \Bigr] ,
\end{eqnarray}
\begin{eqnarray} \label{Lpert}
\lefteqn{\chi^L_{ {\rm pert}} (u) =}
\nonumber \\ & & \frac{1}{8\pi^2 (1-u^2)^3}
\Bigl[ (1-u^2) (1+u+u^2) (1-4u+u^2) - 6u^3 \ln (u^2) \Bigr] \nonumber
\\ & & + \frac{\alpha_s}{48\pi^3 (1-u^2)^4}
\Bigl[ (1-u^2)^2 (1-36u-22u^2-36u^3+u^4) \Bigr. \nonumber \\
& & \hspace{2em} - 2u (1-u^2) (9+4u+66u^2+4u^3+9u^4) \ln (u^2)
\nonumber \\ & &
\hspace{2em} -4u^3 (9+18u^2-2u^3-3u^4+u^5) \ln^2 (u^2) + 8(1-u^2)^3
(1-3u+u^2)
\nonumber \\ & & \Bigl. \hspace{4em} \times {\rm Li}_2 (1-u^2) \Bigr] ,
\end{eqnarray}
where the dilogarithm is defined by
\begin{equation}
{\rm Li}_2 (z) \equiv -\int_0^{z} dz' \, \frac{\ln (1-z')}{z'} .
\end{equation}
Expressions for an axial current $J=A$ are obtained from
(\ref{Tpert}), (\ref{Lpert}) by replacing $u \rightarrow
-u$.

	It has been pointed out\cite{JM} that non-analytic quark mass
dependence, such as in the perturbative results presented above,
indicates the inclusion of some infrared effects into the Wilson
Coefficients (WC's), in conflict with the usual interpretation of the
OPE as a separation into short- and long-distance effects.  A formal
analysis reshuffles the WC's in front of each nonperturbative
condensate.  However, since our analysis requires only the numerical
sum of such effects, the total result should be the same in either
form.

	The leading nonperturbative corrections are supplied by the
condensates of dimension less than five, namely the gluon and quark
condensates.  For a heavy $b$ quark decaying into a quark $q$ of
arbitrary mass through a vector current, the contributions from the
condensates, derived using Refs.~\cite{RRY,GenNP} and evaluated at
$q^2=0$, read
\begin{eqnarray} \label{hl1}
\lefteqn{\chi^T_{\rm cond} (u) =} & & \nonumber \\ & &
- \langle \bar q q \rangle \frac{(2-3u+2u^2)}{2m_b^5 (1-u^2)^5}
\nonumber \\ & &
+ \left< \frac{\alpha_s}{\pi} G^2 \right>
\Biggl\{ \frac{-1}{24 m_b^6 (1-u^2)^7} \Biggr. \nonumber \\ & &
\left. \times \left[ \left( 1-u^2 \right)
\left( 2-104u+148u^2-270u^3+145u^4-104u^5+5u^6-2u^7 \right)
\right. \right. \nonumber \\ & &
\Biggl. \left. \hspace{6em} -12u \ln (u^2) \left(
3-5u+17u^2-15u^3+17u^4-5u^5+3u^6 \right) \right] \Biggr\} , \\
\label{hl2}
\lefteqn{\chi^L_{\rm cond} (u) 
=} & & \nonumber \\ & &
+ \langle \bar q q \rangle \frac{1}{m_b^3 (1-u^2)^3}
\nonumber \\
& & + \left< \frac{\alpha_s}{\pi} G^2 \right>
\Biggl\{ \frac{1}{12m_b^4 (1-u^2)^5} \left[ \left( 1-u^2 \right) \left(
1-21u+10u^2-20u^3+u^4-u^5 \right) \right. \Biggr. \nonumber \\
& &  \Biggl. \left. \hspace{12em} -3u \ln (u^2)
\left( 3-2u+8u^2-2u^3+3u^4 \right) \right] \Biggr\} .
\end{eqnarray}

	These expressions superficially appear to diverge in the limit
$u \to 1$.  However, in this limit the $\langle m_q \bar q q \rangle$
condensate obeys the relation\cite{Witt}
\begin{equation}
\langle m_q \bar q q \rangle = - \frac{1}{12}
\left< \frac{\alpha_s}{\pi} G^2 \right> +
O \left( \frac{1}{m_q^2}, \alpha_s \right) ,
\end{equation}
and Eqs.~(\ref{hl1}), (\ref{hl2}) simplify to 
\begin{eqnarray} \label{hh1}
\chi^T_{\rm cond} (u) & = &
\left< \frac{\alpha_s}{\pi} G^2 \right>
\frac{1}{24 \left( 1-u^2 \right)^7 m_b^6 u} \nonumber \\ & &
\times \Biggl\{ \left( 1-u^2
\right) \left( 2-5u+104u^2-145u^3+268u^4-145u^5+104u^6-5u^7+2u^8
\right) \Biggr. \nonumber \\ & & \Biggl. \hspace{2em} + 12 u^2 \ln
(u^2) \left( 3-5u+17u^2-15u^3+17u^4-5u^5+3u^6 \right) \Biggr\} ,
\end{eqnarray}
\begin{eqnarray} \label{hh2}
\chi^L_{\rm cond}(u) & = & 
-\left< \frac{\alpha_s}{\pi} G^2 \right>
\frac{1}{12 \left( 1-u^2 \right)^5 m_b^4 u}
\Biggl\{ \left( 1-u^2 \right) \left( 1-u+20u^2-10u^3+20u^4-u^5+u^6
\right) \Biggr. \nonumber \\ & & \Biggl. \hspace{13em}
+ 3u^2 \ln (u^2) \left( 3-2u+8u^2-2u^3+3u^4 \right) \Biggr\} ,
\end{eqnarray}

	The expressions (\ref{hh1}), (\ref{hh2}) are applicable to $b
\rightarrow c$ decays, while (\ref{hl1}), (\ref{hl2}) are best suited
to $b \rightarrow u$ or $c \rightarrow s$ transitions.  Expressions
for an axial current are obtained by replacing $u \rightarrow -u$.
The limiting values for $u \rightarrow 0$ are
\begin{eqnarray}
m_b^2 \, \chi^T(0) 
& = & \frac{3}{32\pi^2} + \frac{\alpha_s}{192\pi^3}
(25+4\pi^2) - \frac{1}{m_b^3} \langle \bar q q \rangle -
\frac{1}{12 m_b^4} \left< \frac{\alpha_s}{\pi} G^2 \right> ,
\nonumber \\
\chi^L(0) & =&
\frac{1}{8\pi^2} + \frac{\alpha_s}{144\pi^3} (3+4\pi^2) +
\frac{1}{m_b^3} \langle \bar q q \rangle +
\frac{1}{12m_b^4} \left< \frac{\alpha_s}{\pi} G^2 \right> ,
\end{eqnarray}
while for $u \rightarrow 1$,
\begin{eqnarray}
m_b^2 \chi^T(+1) & =& \frac{1}{20\pi^2} +
\frac{41 \alpha_s}{162\pi^3} -\frac{1}{210 m_b^4}
\left< \frac{\alpha_s}{\pi} G^2 \right> , \nonumber \\
m_b^2 \chi^T(-1) & =& \frac{1}{40\pi^2} +
\frac{689 \alpha_s}{6480\pi^3} -\frac{1}{140 m_b^4}
\left< \frac{\alpha_s}{\pi} G^2 \right> , \nonumber \\
\chi^L(+1) & =& 0 , \nonumber \\
\chi^L(-1) & =& \frac{1}{4\pi^2} +
\frac{7 \alpha_s}{12 \pi^3} +\frac{1}{60 m_b^4}
\left< \frac{\alpha_s}{\pi} G^2 \right> .
\end{eqnarray}
For $b \rightarrow c$, using pole mass values such that $u=0.33$, we
have
\begin{eqnarray}
m_b^2 \chi^T(+0.33) & = & 9.659 \cdot 10^{-3} \left[ 1 + 1.42 \alpha_s
- 4.8 \cdot 10^{-4} \left( \frac{4.9 \, \rm{GeV}}{m_b} \right)^4
\left< \frac{\left( \alpha_s G^2/\pi \right)}{0.02 \, \rm{GeV}^4}
\right> \right] , \nonumber \\
m_b^2 \chi^T(-0.33) & = & 5.709 \cdot 10^{-3} \left[ 1 + 1.32 \alpha_s
- 6.8 \cdot 10^{-4} \left( \frac{4.9 \, \rm{GeV}}{m_b} \right)^4
\left< \frac{\left( \alpha_s G^2/\pi \right)}{0.02 \, \rm{GeV}^4}
\right> \right] , \nonumber \\
\chi^L(+0.33) & = & 3.713 \cdot 10^{-3} \left[ 1 + 1.37 \alpha_s -
5.3 \cdot 10^{-4} \left( \frac{4.9 \, \rm{GeV}}{m_b} \right)^4
\left< \frac{\left( \alpha_s G^2/\pi \right)}{0.02 \, \rm{GeV}^4}
\right> \right] , \nonumber \\
\chi^L(-0.33) & = & 2.162 \cdot 10^{-3} \left[ 1 + 0.64 \alpha_s +
2.2 \cdot 10^{-4} \left( \frac{4.9 \, \rm{GeV}}{m_b} \right)^4
\left< \frac{\left( \alpha_s G^2/\pi \right)}{0.02 \, \rm{GeV}^4}
\right> \right] ,
\end{eqnarray}
where the central value $< \frac{\alpha_s}{\pi} G^2> = 0.02$ GeV${}^4$
is taken from Ref.~\cite{Nar}, and the pole mass value $m_b = 4.9$ GeV
is from \cite{PDG}.  We also use $\alpha_s(m_b) = 0.22$\cite{PDG}, and
since the coefficient of the gluon condensate is tiny, we ignore it in
our numerical analysis for $b \rightarrow c$.

\subsection{Weighting functions \boldmath $\phi$}

        To obtain the general form of the weighting functions $\phi$
defined in Eq.~(\ref{tdisp}), first observe that the quantities
$\Pi^L, \Pi^T$ are, respectively, the $\mu = \nu = 0$ and $\mu = \nu =
1,2$, or 3 components of $\Pi^{\mu\nu}$ evaluated in the center of
mass frame, $q^\mu = ( \sqrt{t},{\bf 0})$ .  Then the generic
expression for the contribution of a particular form factor $F(t)$ to
the polarization tensor may be denoted by
\begin{eqnarray} \label{sat}
{\rm Im} \, \Pi^T &\ge& \frac{n_I}{K \pi} (t-t_+)^{\frac a 2}
(t-t_-)^{\frac b 2} \, t^{-c} | F(t) |^2 \theta (t-t_+) ,
\nonumber \\
{\rm Im} \, \Pi^L &\ge& \frac{n_I}{K \pi} (t-t_+)^{\frac a 2}
(t-t_-)^{\frac b 2} \, t^{-(c+1)} | F(t) |^2 \theta (t-t_+) ,
\end{eqnarray}
where $K$, $a$, $b$, and $c$ are integers determined by the form
factor $F(t)$, and $n_I$ is an isospin Clebsch-Gordan factor, which is
2, 3/2, and 1 for $\bar B \rightarrow D^{(*)}$, $B \rightarrow \pi$,
and $\Lambda_b \rightarrow \Lambda_c$ transitions, respectively.
Also, let $\chi = \chi^T$ or $\chi^L$ denote the generic QCD function
appropriate to the quantum numbers of the form factor.\footnote{
In terms of previous notations, one finds for the meson form factors
considered in Refs.~\cite{BL,BGLNP} $K=2^p \kappa^{-1} / \pi \chi
M^2$, $a = b = p$, $c = s+p-3$, whereas in Ref.~\cite{BS} one finds $K
= 3 \cdot 2^s$, $a = b = w$, and $c = p$.  For the baryon form factors
considered in~\cite{BL}, the relation is given by by $a = 2p+1$, $b =
3-2p$, $c = s+1$, and $K = 2 \kappa^{-1}/\pi \chi M^2 $ or $K = 2
\kappa^{-1}/\pi \chi^L$, depending upon whether the form factor
appears with $\chi$ or $\chi^L$.}
The expressions for the weighting functions are readily derived from
Eq.~(\ref{sat}), and are given by
\begin{eqnarray}\label{phioft}
\phi_i (t;t_0) & = & \sqrt{\frac{n_I}{K \pi \chi}} \,
\left( \frac{t_+ - t}{t_+ - t_0} \right)^{\frac 1 4}
\left( \sqrt{t_+ - t} + \sqrt{t_+ - t_0} \right)
\left( t_+ - t \right)^{\frac a 4} \nonumber \\ & &
\times \left( \sqrt{t_+ - t} + \sqrt{t_+ - t_-} \right)^{\frac b 2}
\left( \sqrt{t_+ - t} + \sqrt{t_+} \right)^{-(c+3)} .
\end{eqnarray}
The values of the parameters $K, a, b, c$ for each form factor, as
well as the relevant $\chi$, are given in Table~1 for $\bar B
\rightarrow D^{(*)}$ transitions, and in Table~2 for $\Lambda_b
\rightarrow \Lambda_c$.  Although (\ref{phioft}) assumes $q^2 =0$, it
is easy to generalize to arbitrary $Q^2 \equiv -q^2$: Simply evaluate
the perturbative functions $\chi(Q^2)$ at the given value, and
multiply $\phi$ by
\begin{equation} \label{phiQ}
{\left( \sqrt{t_+ - t^{\vphantom{2}}} +
\sqrt{t_+^{\vphantom{2}}} \over
\sqrt{t_+ - t^{\vphantom{2}}} + \sqrt{t_+ + Q^2}
  \right)}^{d} ,
\end{equation}
with $d=3$ if the form factor involves $\chi^T$ and $d=2$ if it
involves $\chi^L$.

\bigskip
\vbox{\medskip
\hfil\vbox{\offinterlineskip
\hrule
\halign{&\vrule#&\strut\quad\hfil$#$\quad\cr
height2pt&\omit&&\omit&&\omit&&\omit&&\omit&&\omit&\cr
& F_i &&  K && \chi \quad && a && b && c  & \cr
\noalign{\hrule}
& f_+ && 48 && \chi^T(+u) && 3 && 3 && 2  &\cr
& f_0 && 16 && \chi^L(+u) && 1 && 1 && 1  &\cr
\noalign{\hrule}
& f   && 24 && \chi^T(-u) && 1 && 1 && 1  &\cr
& {\cal F}_1 && 48 && \chi^T(-u) && 1 && 1 && 2  &\cr
& g   && 96 && \chi^T(+u) && 3 && 3 && 1  &\cr
& {\cal F}_2 && 64 && \chi^L(-u) && 3 && 3 && 1 &\cr } \hrule} \hfil
\medskip
\\
\INSERTCAP{1}{Factors entering Eq.~(\ref{phioft}) or (\ref{phiofz})
for the meson form factors $F_i$ in $\bar B \rightarrow D^{(*)}$.}}

\bigskip
\vbox{\medskip
\hfil\vbox{\offinterlineskip
\hrule
\halign{&\vrule#&\strut\quad\hfil$#$\quad\cr
height2pt&\omit&&\omit&&\omit&&\omit&&\omit&&\omit&\cr
& F_i &&  K && \chi \quad && a && b && c & \cr
\noalign{\hrule}
& F_0 &&  8 && \chi^L(+u) && 3 && 1 && 1  &\cr
& F_1 && 12 && \chi^T(+u) && 1 && 3 && 1  &\cr
& H_V && 24 && \chi^T(+u) && 1 && 3 && 2  &\cr
& G_0 &&  8 && \chi^L(-u) && 1 && 3 && 1  &\cr
& G_1 && 12 && \chi^T(-u) && 3 && 1 && 1  &\cr
& H_A && 24 && \chi^T(-u) && 3 && 1 && 2  &\cr } \hrule} \hfil
\medskip
\\
\INSERTCAP{2}{Factors entering Eq.~(\ref{phioft}) or (\ref{phiofz})
for the baryon form factors $F_i$.}}

	While the momentum-transfer variable $t = (p-p')^2$ is useful
for heavy-to-light decays and has an obvious physical meaning, it is
often more convenient when dealing with heavy-to-heavy transitions
such as $ b\to c$ to use a kinematic variable that helps disentangle
long-distance physics from the heavy quark scale.  One such variable
is
\begin{equation}\label{wdef}
w \equiv v \cdot v' = \frac{M^2+m^2-t}{2Mm} .
\end{equation}
In the $b$ rest frame, $w$ depends only on the energy transfer to the
light degrees of freedom, in units of $\Lambda_{\rm QCD}$.  It is due
to this property that $b \to c$ form factors are related to each other
in the heavy quark limit at equal values of $w$. In the semileptonic
region, the variable $z(t;t_0)$ has the same physical property, so
form factors will be related at equal values of $z$ as well. We can
demonstrate this by expressing $z$ as a function of $w$ without
reference to heavy meson masses,
\begin{equation} \label{zofw}
z(w;N) = { \sqrt{ 1+w} - \sqrt{2 N} \over 
         \sqrt{ 1+w} + \sqrt{2 N} } \ ,
\end{equation}
where $N$ is a free parameter related to $t_0$ by
\begin{equation}\label{Ndef}
 N = { t_+ -t_0 \over t_+ -t_-} ,
\end{equation}
so that $z(w; N)$ vanishes at $w = 2 N -1$. With $t, t_0$ related to
$w,N$ by Eqs.~(\ref{wdef}) and (\ref{Ndef}), Eq.~(\ref{zofw}) is
simply a rewriting of Eq.~(\ref{zblaschke}), $z(t;t_0) = z(w;N)$. The
advantage in using $z(w;N)$ for $b \to c$ transitions is that its
definition is process-independent.

	For the semileptonic decay $H_b \to H_c \ell \bar \nu$, the
limiting values of $z$ are given by $m_\ell^2 \leq t \leq t_-$, or
\begin{equation}
z_{\rm min} = - \left( \frac{\sqrt{N}-1}{\sqrt{N}+1} \right)
\end{equation}
and
\begin{equation}
z_{\rm max} = \frac{\sqrt{(1+r)^2-\delta^2} -
2\sqrt{Nr}}{\sqrt{(1+r)^2-\delta^2} + 2\sqrt{Nr}} \  .
\end{equation}
where $r = m/M$ and $\delta = m_\ell/M$.

	The dispersion relation (\ref{tdisp}), written now entirely in
terms of $z$, reads
\begin{equation} \label{dispz}
\frac{1}{2\pi i} \int_C \frac{dz}{z} |\phi(z) P(z) F(z)|^2 \leq 1 ,
\end{equation}
where $C$ is the unit circle in the complex $z$ plane, the Blaschke
factor for a pole at $z_p \equiv z(t_p; t_0)$ (which is real for
sub-threshold resonance masses) is
\begin{equation}
z(t; t_p)  = \frac{z-z_p}{1-zz_p},
\end{equation}
for $z = z(t; t_0)$ and any $t_0$, with $P(z)$ being the product of
all such factors, and the weighting functions (\ref{phioft}) are given
by
\begin{eqnarray} \label{phiofz}
\phi (z;N) & = & M^{\frac 1 2 (a+b) - (c+2)} \, \sqrt{\frac{n_I}{K
\pi \chi}} \, 2^{\frac 1 2 (a+b) +2} \, N^{\frac a 4 + \frac 1 2} \,
r^{\frac 1 4 (a+b) + \frac 1 2} \, (1+z)^{\frac 1 2 (a+1)}
(1-z)^{c +  \frac 1 2 (3-a-b)} \nonumber \\ & &
\times \left[ \left( \sqrt{N}-1 \right) z
+ \left( \sqrt{N}+1 \right) \right]^{\frac b 2}
\left[ (1+r)(1-z) + 2 \sqrt{Nr} (1+z) \right]^{-(c+3)}  .
\end{eqnarray}
Finally, evaluation of $\phi (z;N)$ at nonzero values of $Q^2 = -q^2$
is accomplished by multiplying (\ref{phiofz}) by
\begin{equation}
\left( \frac{(1+r)(1-z) + 2\sqrt{Nr}(1+z)}{\sqrt{(1+r)^2 +
Q^2/M^2} (1-z) + 2 \sqrt{Nr} (1+z)} \right)^d,
\end{equation}
where $d$ is defined as in Eq.~(\ref{phiQ}).

\subsection{Sub-threshold Resonances and \boldmath $P(t)$}

        The Blaschke factor $P(t)$ for a form factor describing $H_b
\to H_c \ell \bar \nu$ depends on the masses of $B_c$ resonances below
the $H_b \bar H_c$ pair-production threshold.  The Blaschke factors
are simply products of $z(t;t_s)$ with $t_s$ evaluated at the
invariant mass-squared $t_s \rightarrow t_p$ of each such resonance
with the same spin-parity as the current $J$,
\begin{equation} \label{Poft}
P(t) = \prod_p z(t;t_p) .
\end{equation}
For $b \to c $ transition from factors, the masses of the relevant
$B_c$-type resonances can be accurately estimated from potential
models\cite{pot,Quigg}.  We compile in Table~3 the masses computed in
Ref.~\cite{Quigg}.

\bigskip
\vbox{\medskip
\hfil\vbox{\offinterlineskip
\hrule
\halign{&\vrule#&\strut\quad\hfil$#$\quad\cr
height2pt&\omit&&\omit&\cr
& {\rm Type} &&  {\rm Masses} \; {\rm (GeV)} \qquad
& \cr \noalign{\hrule}
& {\rm Vector} && 6.337, \; 6.899, \; 7.012, \;
7.280 &\cr
&  && 7.350, \; 7.594, \; 7.646, \; 7.872, \; 7.913 &\cr
\noalign{\hrule}
& {\rm Axial} \; {\rm vector}  && 6.730, \; 6.736, \; 7.135, \; 7.142
& \cr
&
&& 7.470, \; 7.470, \; 7.757, \; 7.757 &\cr
\noalign{\hrule}
& {\rm Scalar}  && 6.700, \; 7.108, \; 7.470, \; 7.757 \; &\cr
\noalign{\hrule}
& {\rm Pseudoscalar} && 6.264, \; 6.856, \;
7.244, \; \; 7.562, \; 7.844 \;  &\cr
}
\hrule}
\hfil
\medskip
\\
\INSERTCAP{3}{Calculated $B_c$ pole masses used in this work.}}

	For heavy-to-light form factors there is no formal limit in
which the light quark becomes nonrelativistic, and potential model
calculations are less reliable. This is not a problem for $B \to \pi$,
where the only sub-threshold resonance, the $B^*$, is experimentally
observed, but for decays to other light states such as $B \to \rho$,
the presence and masses of additional sub-threshold resonances must be
taken from models.  Once these uncertainties are accounted for, simple
parameterizations should be reliable. For example, the model of
\cite{ehq}\ indicates that the form factor $f$ for $B \to \rho$ has
only one narrow sub-threshold pole.  Indeed, this pole appears to have
been observed (mixed with others) by ALEPH, DELPHI, and OPAL [see
\cite{PDG} for analysis and references on the ``$B^*_J$(5732)''].
Even accounting for significant uncertainties in its mass, this leads
to an accurate parameterization using the overall normalization and
two unknown coefficients.  It is important to estimate the
uncertainties from model-dependent poles on a case by case
basis\cite{Bec2}.

	This applies as well to sub-threshold branch cuts due
to multi-particle states and anomalous thresholds. A model-dependent
analysis\cite{BL,BGLNP} suggests these are negligible for $b \to c$
transitions.  Qualitatively, this result comes about because cuts are
a much less severe form of non-analytic behavior than poles.  Whether
cuts continue to be unimportant for $B \to \rho$ transitions requires a
more detailed analysis.

\section{Additional States} \label{add}

	The effects of higher states in the dispersion relation depend
on the flavor of the $b \to q$ current under consideration.
Henceforth, we specialize to $b \to c$ transitions, for which HQET is
most useful.

\subsection{Contributions to the Dispersion Relation}

	We have observed that contributions to the original dispersion
relation [see Eq.~(\ref{ins})] are semipositive definite, so each
additional state coupling to the vacuum through the current $J$ serves
to further saturate the bound supplied by the QCD parton-level
calculation.  The inclusion of only a single two-particle $\bar B \bar
D$ or $\bar B \bar D^{*}$ state in obtaining these bounds is
relatively weak, since such exclusive states account for only a small
portion of the inclusive total.  In general, the hadronic side
includes also $B_c$ resonances, a continuum of states like $\bar B
\bar D \pi \pi$, and so on.  While it is desirable to include as many
of these states as possible, it is not clear how to include them in a
model-independent fashion; the chief exceptions are two-particle
states related to one another via heavy quark spin symmetry, namely
the four states $\bar B^{(*)} \bar D^{(*)}$.  While form factors for
transitions such as $\bar B^* \rightarrow D^{(*)}$ are not physically
accessible through semileptonic decays, their normalization is
nonetheless known via HQET, allowing an additional strengthening of
the dispersive bounds.

	The contributions (\ref{sat}) of these states to the
dispersion relations are of the form
\begin{eqnarray}
{\rm Im} \, \Pi(t) &\ge& \sum_i \kappa_i(t) |F_i(t)|^2 ,
\end{eqnarray}
where the sum is over all helicity amplitudes $F_i$ arising from pair
production of $\bar B^{(*)} \bar D^{(*)}$. The weight functions $\phi$
are readily obtained from the kinematic prefactors $\kappa_i(t)$,
while the relation between the helicity amplitudes and the original
form factors may be obtained by choosing definite polarizations of the
$\bar B^*$, $\bar D^*$, and the virtual $W$.

	For $\bar B^* \rightarrow D$ transitions, the form factors are
defined by
\begin{eqnarray}
\langle D (p') | V^\mu | \bar B^* (p,\epsilon) \rangle & = & i \hat g
\epsilon^{\mu \alpha \beta \gamma} \epsilon_\alpha p_\beta p'_\gamma
, \nonumber \\
\langle D^*(p') | A^\mu | \bar B^* (p,\epsilon) \rangle & = & \hat f
\epsilon^\mu + (\epsilon \cdot p') \left[ \hat a_+ (p'+p)^\mu + \hat
a_- (p'-p)^\mu \right] .
\end{eqnarray}
	The functions $\phi$ for $\bar B^* \rightarrow D$ transitions
are identical to those for $\bar B \rightarrow D^*$, with the simple
replacement $M \leftrightarrow m$.  The helicity amplitudes possess
factors of $M$ and $m$, and need not be invariant under this exchange.
This is true in particular for ${\cal F}_1$ and ${\cal F}_2$; here we
find that [compare (\ref{F12})]
\begin{eqnarray}
\hat {\cal F}_1 (t) & \equiv & \frac 1 M \Biggl[ 2k^2 t \hat a_+ (t)
- \frac 1 2 (t-m^2+M^2) \hat f (t) \Biggr] , \nonumber \\
\hat {\cal F}_2 (t) & \equiv & \frac 1 M \Biggl[ \hat f (t) - (M^2-m^2)
\hat a_+ (t) + t \hat a_- (t) \Biggr] .
\end{eqnarray}

	The transition $\bar B^* \rightarrow D^*$ possesses 10
independent vector current form factors, which we define as
follows\cite{BGnet}:
\begin{eqnarray}
\frac{1}{\sqrt{Mm}} \langle D^* (\epsilon_D , v') | V^\mu | \bar B^*
(\epsilon_B , v) \rangle & = & f_4 (\epsilon_B \cdot \epsilon_D^*)
v'^\mu + f_5 (\epsilon_B \cdot \epsilon_D^*) v^\mu \nonumber \\ & & +
f_6 (v' \cdot \epsilon_B) (v \cdot \epsilon_D^*) v^\mu + f_7 (v' \cdot
\epsilon_B) (v \cdot \epsilon_D^*) v'^\mu \nonumber \\ & &
+ f_8 (v \cdot \epsilon_D^*) \epsilon_B^\mu + f_9 (v' \cdot
\epsilon_B) \epsilon_D^{*\mu} , \nonumber \\
\frac{1}{\sqrt{Mm}} \langle D^* (\epsilon_D , v') | A^\mu | \bar B^*
(\epsilon_B , v) \rangle & = & 
+ i f_{10} \epsilon^\mu{}_{\alpha\beta\gamma} \epsilon_B^\alpha
\epsilon_D^{*\beta} v'^\gamma + i f_{11}
\epsilon^\mu{}_{\alpha\beta\gamma} \epsilon_B^\alpha
\epsilon_D^{*\beta} v^\gamma \nonumber \\ & &
+ i f_{12} \left[ (v' \cdot \epsilon_B)
\epsilon^\mu{}_{\alpha\beta\gamma} \epsilon_D^{*\alpha} v'^\beta
v^\gamma + \epsilon_{\alpha\beta\gamma\delta} \epsilon_B^{\alpha}
\epsilon_D^{*\beta} v'^\gamma v^\delta v'^\mu \right] \nonumber \\ & &
+ i f_{13} \left[ (v \cdot \epsilon_D)
\epsilon^\mu{}_{\alpha\beta\gamma} \epsilon_B^{*\alpha} v'^\beta
v^\gamma - \epsilon_{\alpha\beta\gamma\delta} \epsilon_B^{\alpha}
\epsilon_D^{*\beta} v'^\gamma v^\delta v^\mu \right] .
\end{eqnarray}
The combinations of these form factors appearing as helicity
amplitudes may be denoted
\begin{eqnarray}
V_{++} & = & \frac{f_9}{\sqrt{Mm}} ,\nonumber \\
V_{+0} & = & \frac{f_8}{\sqrt{Mm}} ,\nonumber \\
V_{0+} & = & \frac{1}{\sqrt{2Mm}} \left( Mf_4 + mf_5 \right) ,
\nonumber \\
V_{00} & = & \frac{1}{4(Mm)^{\frac52}} \Bigl\{ Mm \left(
t-M^2-m^2 \right) \left( Mf_4 + mf_5 \right) - 2k^2 t \left( mf_6
+ Mf_7 \right) \nonumber \Bigr. \\ & & \Bigl. + Mm \left[ m \left(
t + M^2 - m^2 \right) f_8 + M
\left( t - M^2 + m^2 \right) f_9 \right] \Bigr\} , \nonumber \\
S_{00} & = & \frac{1}{(2Mm)^{\frac52}} \Bigl\{ Mm
\left(t-M^2-m^2\right) \left[ M \left(t-M^2+m^2 \right) f_4 - m
\left( t + M^2 - m^2 \right) f_5 \right] \Bigr. \nonumber \\ & &
\Bigl. + 2 k^2 t \left[ m \left( t+M^2-m^2 \right) f_6 - M
\left( t - M^2 + m^2 \right) f_7 \right] - 4k^2 t Mm \left( mf_8
- Mf_9 \right) \Bigr\} , \nonumber \\
S_{0+} & = & \frac{1}{\sqrt{4Mm}} \left[ M \left( t - M^2 + m^2
\right) f_4 - m \left( t + M^2 - m^2 \right) f_5 \right] , \nonumber
\\
A_{++} & = & \frac{1}{2(Mm)^{\frac32}} \left\{ Mm \left[ \left( t
- M^2 - m^2 \right) f_{10} - 2Mm f_{11} \right] - 2k^2 t f_{12}
\right\} , \nonumber \\
A_{0+} & = & \frac{1}{2(Mm)^{\frac32}} \left\{ M^2 m \left( t -
M^2 + m^2 \right) f_{10} - Mm^2 \left( t + M^2 - m ^2 \right) f_{11}
\right. \nonumber \\ & & \left. + 2k^2
t \left( Mf_{12} - mf_{13} \right) \right\} , \nonumber \\
A_{+0} & = & \frac{1}{2(Mm)^{\frac32}} \left\{ Mm \left[ 2Mm f_{10}
- \left( t-M^2-m^2 \right) f_{11} \right] + 2k^2 t f_{13} \right\}
, \nonumber \\
P_{0+} & = & \frac{1}{2(Mm)^{\frac32}} \left\{ 2Mm \left( Mf_{10} \!
+ \! mf_{11} \right) + \left[ M \left( t \! - M^2 \! + m^2 \right)
f_{12} + m \left( t \! + M^2 \! - m^2 \right) f_{13} \right]
\right\} . \nonumber \\ & & 
\end{eqnarray}
The labels $V,S,A,P$ reflect the spin-parity (vector, scalar, axial
vector, or pseudoscalar) of the virtual $W$, while the subscripts
denote the helicities of the $W^*$ and $D^*$ in the decay of the $\bar
B^*$.

	Only amplitudes of a fixed spin-parity enter each dispersion
relation. For example, the four $V$ helicity amplitudes enter the vector
current dispersion relation for $\Pi^T$. Above all the appropriate
pair-production thresholds, the contribution to ${\rm Im} \, \Pi^T_V$
from $\bar B \bar D$, $\bar B^* \bar D$, $\bar B \bar D^*$, $\bar B^*
\bar D^*$, and $\Lambda_b \bar \Lambda_c$ states is
\begin{eqnarray} \label{piTVexample}
{\rm Im} \, \Pi^T_V &\ge& { k^3 \over 6 \pi \sqrt{t} } \Biggl[ 2
     \left( |f_+|^2 + |V_{0+}|^2 + |V_{00}|^2 \right) + t \left( |g|^2
     + |\hat g|^2 + |V_{++}|^2 + |V_{+0}|^2 \right) \Biggr]
\nonumber \\
    &+&  { k (t- t_-) \over 12 \pi t^{3/2} }
   \Biggl[ 2 t |F_1|^2 + |H_V|^2 \Biggr] ,
\end{eqnarray}
where we have included an isospin factor $n_I = 2$ for the mesons, and
$k$ is defined in Eq.~(\ref{kdef}). Using Eq.~(\ref{sat}), it is
straightforward to compute the weighting functions $\phi$ for all the
$\bar B^* \bar D^*$ form factors. For each such $F_i$, one obtains a
parameterization of the form Eq.~(\ref{masterparam}), with the unknown
expansion coefficients denoted by $b_{in}$,
\begin{equation}\label{Fparam}
F_i(z) = {1 \over P_i(z) \phi_i(z) } \sum_{n=0}^\infty
                   b_{in} z^n .
\end{equation}
The $\phi$ parameters [see (\ref{phioft}) or (\ref{phiofz})] $a,b,c,K$
and the relevant $\chi$ for $\bar B^* \rightarrow D^*$ are given in
Table~4.

\bigskip
\vbox{\medskip
\hfil\vbox{\offinterlineskip
\hrule
\halign{&\vrule#&\strut\quad\hfil$#$\quad\cr
height2pt&\omit&&\omit&&\omit&&\omit&&\omit&&\omit&\cr
& F_i &&  K && \chi \quad && a && b && c  & \cr
\noalign{\hrule}
& V_{++} && 96 && \chi^T(+u) && 3 && 3 && 1  &\cr
& V_{+0} && 96 && \chi^T(+u) && 3 && 3 && 1  &\cr
& V_{0+} && 48 && \chi^T(+u) && 3 && 3 && 2  &\cr
& V_{00} && 48 && \chi^T(+u) && 3 && 3 && 2  &\cr
& S_{00} &&  8 && \chi^L(+u) && 1 && 1 && 1  &\cr
& S_{0+} &&  8 && \chi^L(+u) && 1 && 1 && 1  &\cr
& A_{++} && 24 && \chi^T(-u) && 1 && 1 && 1  &\cr
& A_{0+} && 24 && \chi^T(-u) && 1 && 1 && 2  &\cr
& A_{+0} && 24 && \chi^T(-u) && 1 && 1 && 1  &\cr
& P_{0+} && 32 && \chi^L(-u) && 3 && 3 && 1  &\cr} \hrule} \hfil
\medskip
\\
\INSERTCAP{4}{Factors entering Eq.~(\ref{phioft}) or (\ref{phiofz})
for the meson form factors $F_i$ in $\bar B^* \rightarrow D^*$.}}

	Substituting this expansion into the dispersion relation
(\ref{dispz}) gives
\begin{equation} \label{allbd}
\sum_{i=0}^{H} \sum_{n=0}^{\infty} b_{in}^2 \le 1.
\end{equation}
Included in the sum are all helicity amplitudes $i=0,\cdots,H$ for
processes with the right quantum numbers to couple to the current
$J$. It is clear that the constraint on a particular helicity
amplitude $F_i$ can be strengthened if it is possible to relate the
various $b_{in}$, {\it i.e.}, if one can relate the form factors. This
is accomplished with the help of heavy quark symmetry.

\subsection{Form factors in the Heavy Quark Limit}

	In general, form factors are not related by heavy quark
symmetry throughout the pair production region\cite{Ball}, but may be 
related in the semileptonic region.  This fact has been exploited to
improve constraints on the $B \to B$ elastic form factor\cite{CM} and
the dominant $\bar B \to \pi \ell \bar \nu$ form factor\cite{BS}.  The
situation in the present case is conceptually analogous, although
algebraically more cumbersome.
 
	In the semileptonic region, the 20 form factors of $\bar
B^{(*)} \rightarrow D^{(*)}$ reduce to only one in the heavy quark
limit, the universal Isgur-Wise function $\xi (w)$, with $\xi (1) =
1$.  Likewise, the six baryon form factors in $\Lambda_b \rightarrow
\Lambda_c$ reduce to another universal function\cite{IWbary}, which we
may denote $\zeta (w)$, with $\zeta (1) = 1$.  Recalling that $r =
m/M$, the relation of the helicity amplitudes to $\xi$ and $\zeta$ in
the heavy quark limit are given by
\begin{eqnarray} \label{mesonhqs}
 f_+ & = & \frac12 {\cal F}_2 = -\frac12 \hat{\cal F}_2 =
-\frac1{\sqrt{2}} V_{0+} = V_{00} = \frac12 P_{0+} =
\frac{(1+r)}{2\sqrt{r}}    \xi , \nonumber \\
f_0 & = & {\cal F}_1 = \hat{\cal F}_1 = -\sqrt{2} S_{00}
      = S_{0+} = - A_{0+} =
M^2 \sqrt{r} (1-r) (1+w) \xi , \nonumber \\
g   & = & -\hat{g} = V_{++} =  V_{+0} = 
\frac{1}{M\sqrt{r}}      \xi , \nonumber \\
f   & = & -\hat{f} = -A_{++} = A_{+0} =
M \sqrt{r} (1+w)         \xi , 
\end{eqnarray}
for mesons, and by
\begin{eqnarray} \label{baryonhqs}
F_0 & = & H_A = M(1-r) \zeta , \nonumber \\
F_1 & = & G_1 =        \zeta , \nonumber \\
H_V & = & G_0 = M(1+r) \zeta   
\end{eqnarray}
for baryons.  The only helicity amplitudes in this list protected by
Luke's theorem\cite{Luke,BBLS} from $1/M$ corrections at $w=1$ are
$f$, $\hat f$, and $G_1$.

	In the strict heavy quark limit, our dispersion relation
constraints become useless because the form factors in
Eqs.~(\ref{mesonhqs}) and (\ref{baryonhqs}) develop an essential
singularity due to an infinite number of poles just below
threshold\cite{spoilers}.  The description Eq.~(\ref{masterparam})
then contains no information, because the Blaschke factor $P(t)$ goes
to zero in the semileptonic region.

	For finite masses our parameterizations are well-behaved, and
heavy quark relations are valid up to $1/m$ corrections.  In this
case, some of the form factors in Eqs.~(\ref{mesonhqs}) or
(\ref{baryonhqs}) will have more constraining parameterizations than
others because their Blaschke factors, which reflect the number and
positions of sub-threshold resonances, will be larger.  If one views
the universality of the Isgur-Wise function as arising from the
dominance of the essential singularity in each of the form factors,
one might expect $1/m$ corrections to be larger for form factors with
fewer sub-threshold poles.  Of course, it is always possible that the
residues of the various poles could conspire to keep $1/m$ corrections
small, since Blaschke factors alone permit\cite{Taron} any residues
consistent with the dispersion relation bounds.

	We use the heavy quark relations (\ref{mesonhqs}) and
(\ref{baryonhqs}) to tighten our bounds on parameterization
coefficients. When we use these bounds to quote smaller errors on our
parameterizations, we allow for substantial deviations due to $1/m$
effects, thereby minimizing errors induced by assuming full heavy
quark symmetry. However, when we use these bounds to test heavy quark
symmetry by constraining the slope or curvature of the Isgur-Wise
function, one should bear in mind the possibility that heavy quark
violations could be larger in form factors with very few sub-threshold
poles (such as $f_0, S_{00}$, or $S_{0+}$) than in those that are
typically measured experimentally.

\subsection{Bounding Parameterization Coefficients}

	We now use the heavy quark relations of the higher resonance
helicity amplitudes to improve the constraints. For concreteness,
consider the form factor $g(w)$. Near zero recoil, $w=1$, heavy quark
symmetry relates it to six other form factors appearing in the
dispersion relation,
\begin{eqnarray} \label{hqsg}
g(w) &=& -\hat g(w) =  V_{++} = V_{+0} \nonumber \\
     &=& \frac2{M+m} f_+(w) 
      = \frac2{M+m} V_{00} = -\frac{\sqrt{2}}{M+m} V_{0+}.
\end{eqnarray}
While exact heavy quark spin symmetry implies that the functional
dependence of these form factors is the same for all values of $w$
corresponding to semileptonic decay, for physical masses we demand
only that the normalization and first and second derivatives of these
form factors are roughly equal at $w=1$.  This suffices to provide
lower bounds, assuming full heavy quark symmetry, on the contributions
of the form factors in Eq.~(\ref{hqsg}) to the sum in
Eq.~(\ref{allbd}).  Once computed, it is straightforward to weaken
these lower bounds by including factors indicating the violation of
heavy quark symmetry.

Since each form factor
$F_i$ in (\ref{hqsg}) has a parameterization of the form
(\ref{Fparam}), the coefficients $a_n$ in the expansion of $g$ can be
related to the coefficients $b_{in}$ in the expansion of $F_i$ by
\begin{equation}\label{hqsC}
\sum_n a_n z^n \approx \frac1{C_i(z)} \sum_n b_{in} z^n,
\end{equation}
where $\approx$ means only the normalization and first two derivatives
are equal at $w=1$. The functions
\begin{equation}
C_i = {\Xi_g P_i \phi_i \over \Xi_i P_g \phi_g}
\end{equation}
are given by ratios of Blaschke factors, weighting functions, and
symmetry factors $\Xi_i(z)$ chosen so that
\begin{equation}
\xi(z) = \Xi_i(z) F_i(z)
\end{equation}
in the heavy quark limit.  The kinematic factors $\Xi_i(w)$ appear in
Eqs.~(\ref{mesonhqs}) and (\ref{baryonhqs}).  By choosing the same
value of $N $ for each form factor, we ensure that the kinematic
variable $z(w; N)$ is process-independent, most of the $z$ dependence
in $\phi_g(z)/\phi_i(z)$ cancels out, and the $C_i(z)$ become quite
simple.

	Numerically, the values of $N$ that optimize our constraints
correspond to $N = 1 + 2 \epsilon$ with $\epsilon \approx 0.05$, and
ignoring terms of order $\epsilon^2$ in Eq.~(\ref{hqsC}) is a good
approximation for parameter values $a_n$ that saturate their
bounds. Evaluating (\ref{hqsC}) and its first and second derivatives
at $w=1$ ($z \approx -\epsilon/2$), we find for each $i$,
\begin{eqnarray}\label{bsol}
b_{0} &=& C a_0
\nonumber \\
b_{1} &=&  C' a_0 + C a_1
\nonumber \\
b_{2} &=&  \frac12 C'' a_0 + C' a_1 + C a_2 
   + \frac32 \epsilon ( b_3 - b_3^{\rm HQS} ) \ ,
\end{eqnarray}
where $C$, $C' = dC/dz$ and $C'' = d^2C/dz^2$ are evaluated at $z=0$
({\em not} $w=1$), and $b_3^{\rm HQS}$ is the value $b_3$ would have
if the third derivative of (\ref{hqsC}) yielded a valid relation.
Departures from the heavy quark symmetry limit tend to increase as we
take higher derivatives of (\ref{hqsC}), so one might expect
substantial corrections to $b_3 = b_3^{\rm HQS}$.  However,
corrections to this relation are multiplied by $\epsilon$, so we may
justifiably ignore the factor $3\epsilon ( b_3 - b_3^{\rm HQS})/2$ in
$b_2$.

	Substituting Eq.~(\ref{bsol}) into
\begin{equation} \label{mult}
\sum_{n=0}^\infty a_n^2 +
   \sum_{i=1}^H \sum_{n=0}^\infty b_{in}^2 \le 1
\end{equation}
gives more stringent bounds on the coefficients $a_n$.  For the form
factor $g$, they yield $-0.37 \le a_1 \le 0.40$ and $-0.49 \le a_2 \le
0.47$, compared to $|a_1|, |a_2| \le 1$ from Eq.~(\ref{abound}).
Bounds for the other form factor coefficients can be obtained in a
straightforward fashion by singling out a different set of $b_{in}$ in
(\ref{hqsC}).

	The bounds on $a_1$ are useful as tests of heavy quark
symmetry. For example, parameterizing $f_0$ and constraining $-0.62
\le a_1 \le 0.58$ restricts the slope of the Isgur-Wise function, up
to $1/m_c$ effects, to $-0.3 \le \rho^2 \equiv -(d\xi/dw) |_{w=1} \le
1.8$.  Bounds on $a_2$ are useful for decreasing truncation errors of
our one-coefficient parameterizations, as described in the next
section. In this case, however, we do not want to rely on exact heavy
quark symmetry, because we wish to be as conservative as possible with
how much the symmetry improves the truncation error, so we allow for
explicit violations to the infinite
mass limit.  Such violations are potentially largest for the bounds on
$a_2$, because they depend on relations involving second
derivatives. Allowing for the total contribution from the higher-spin
states to be as low as $60\%$ of their infinite-mass value gives
bounds on $a_2$ of $-0.58 \le a_2 \le 0.57$ for $g$, which leads to a
corresponding decrease in the error induced by truncating the
expansion (\ref{masterparam}) by $40\%$ [see Eq.~(\ref{trunc}) below].
In summary, bounds on $a_1$ are obtained using full, unbroken heavy
quark symmetry, and may be used to test the accuracy of this
symmetry. The bounds on $a_1$ do not enter the construction of our
parameterizations. The derivation of the bounds on $a_2$ allows for
substantial violations of heavy quark symmetry. These bounds enter
into the truncation errors we quote on our parameterizations.

	Since the baryonic and mesonic form factors are not related by
heavy quark symmetry, Eq.~(\ref{mult}) only applies in the baryonic
case to the pairs $F_1, H_V$ and $G_1, H_A$. Of these, only $H_V$ and
$H_A$ receive substantial improvements to their truncation errors,
which are proportional to $a_3$ [see Eq.~(\ref{trunc})].  While
$1/m_c$ corrections to relations involving $a_3$ and $b_3$ could {\it
\`{a} priori\/} be large, they are known in terms of one constant
$\bar \Lambda_\Lambda$ for the baryons\cite{GGW}, and are not
particularly large.  For example, the relation between $H_V$ and $F_1$
is independent of $w$, to $O(1/m_c^2)$.  Allowing for the
contributions to the dispersion relation from $F_1$ and $G_1$ to be as
small as $50\%$ of their heavy quark symmetry values gives $|a_3|_{\rm
max} = 0.57$ for $H_V$ and $|a_3|_{\rm max} = 0.29$ for $H_A$.

	The bounds on $a_1$ ignoring heavy quark violation, and the
bounds on $a_2$ allowing for deviations from the heavy quark limit as
described above, are given in Tables~5 and 6.
 
\section{Truncation Errors} \label{errs}

	To fit data with our parameterizations, we must truncate the
series (\ref{masterparam}) after a finite number $Q$ of unknown
coefficients $a_n$. This introduces a truncation error which can be
minimized by choosing an appropriate value of $N$\cite{BL} (or
equivalently, $t_0$). Rather than use $a_0$ as a free parameter as
in~\cite{BGLNP}, we solve for $a_0$ so that the form factor is
automatically normalized at $w=1$ to $F(1)$.  If $N=1$, then
$z(1;N)=0$, and this parameterization coincides with that
in~\cite{BGLNP}.  For $N \geq 1$, the parameterization
(\ref{masterparam}), including the solution for $a_0$, may be
re-expressed as
\begin{equation}\label{xifexpansion}
\Xi_F(w) F(w) = {\Xi_F(w) \over P(w) \phi(w;N)  }
         \left\{ {P(1) \phi(1;N) F(1) }
   +\sum_{n=1}^{Q} a_n \left[z^n(w;N) - z^n(1;N) \right] \right\} ,
\end{equation}
where the product $\Xi_F F$ is normalized to coincide with the
Isgur-Wise function $\xi(w)$ [or $\zeta(w)$ for baryons] in the heavy
quark limit. The full form factor is of course given by the $Q
\rightarrow \infty$ sum, while approximations $F^{\rm fit}$ are
obtained by truncating at finite $Q$; then the fit coefficients $a_1,
\ldots , a_Q$ can be chosen so that the difference between the
parameterization $F^{\rm fit}$ and the actual form factor $F$ is given
by
\begin{eqnarray} \label{truncpre}
\Xi_F(w) \left( F(w) - F^{\rm fit}(w) \right)  & = &  \frac{\Xi_F(w)}
{P(w)\phi(w) }
\sum_{n=Q+1}^\infty a_n (z^n-z^n_{\min})  \nonumber \\
&\approx& \frac{\Xi_F(w)} {P(w)\phi(w) }
            a_{Q+1} (z^{Q+1}-z^{Q+1}_{\min}),
\end{eqnarray}
where we have ignored numerically unimportant higher order terms.
Using the Schwarz inequality, the boundedness condition $\sum_n a_n^2
\leq 1$, and the geometric series sum, these higher order terms can 
be shown to be smaller than
\begin{eqnarray}
 \left| \frac{\Xi_F(w)}{P(w)\phi(w)} 
\sum_{n=Q+2}^\infty a_n (z^n-z^n_{\min}) \right| 
&\le&  \left| {\Xi_F(w) \over P(w)\phi(w)} \right|
\sqrt{\sum_{n=Q+2}^\infty |a_n|^2 
\sum_{n=Q+2}^\infty \left( z^n - z_{\rm min}^n \right)^2} 
\nonumber \\ 
  &\le &  \left|\frac{\Xi_F(w)}{P(w)\phi(w)} 
        \Bigl( z^{Q+2}-z^{Q+2}_{\min}\Bigr) \right| + O(z^{Q+4})  ,
\end{eqnarray}
which, since $ |z| \leq 0.04$ for the cases at hand, can be
justifiably ignored. Thus, to good accuracy, the maximal possible
truncation error $T(\Xi_F F)$ for any physically allowed $z$ is
\begin{eqnarray} \label{trunc}
T (\Xi_F F) &=& 
  \max \left| {\Xi_F(w)  (z^{Q+1}-z^{Q+1}_{\min}) \over
           P(w)\phi(w) } \right|\ \cdot |a_{Q+1}|_{\rm max},
\end{eqnarray}
where $Q$ is the highest power of $z$ used in the parameterization fit
and $|a_{Q+1}|_{\rm max}$ is the maximal allowed value of $|a_{Q+1}|$.
The inclusion of higher states leads to tighter bounds on
$|a_{Q+1}|_{\rm max}$ and thus smaller truncation errors.

	The truncation error vanishes at $w=1$, where the
normalization is fixed.  For optimal values of $N$, $z_{\rm min}
\approx - z_{\rm max}$, so for one-coefficient parameterizations the
truncation error drops sharply for some $w$ near $w_{\rm max}$.  This
leads to significantly smaller truncation errors than in previous
works\cite{BL,BGLNP}.  For two-coefficient parameterizations, the
factor $(z^3-z^3_{min})$ adds constructively near $w_{max}$; for a
given bound on $|a_3|$, this leads to a larger truncation error than
in previous work\cite{BL}.  This is unavoidable as long as only $a_1$
and $a_2$ are fit parameters, {\it i.e.}, as long as $a_0$ is chosen
to enforce the normalization at zero recoil, so that the truncation
error vanishes at $z=z_{\rm min}$ rather than $z=0$.

	Our truncation errors for one- and two-coefficient
parameterizations of the various form factors in \btod, \btods, and
$\Lambda_b \to \Lambda_c \ell \bar \nu$ are shown in Tables~5 and
6. The value of the free parameter $N$ has been optimized for each
form factor and number of fit parameters to produce the smallest
truncation errors. The bounds on $a_1$ come from Eq.~(\ref{mult}) and
the heavy quark symmetry relation Eq.~(\ref{bsol}), and may be used as
tests of heavy quark symmetry. The bounds on $a_2$ enter our
truncation errors, so allowance for heavy quark symmetry violation has
been made as described in the previous section. The truncation error
is expressed as a percentage of $\Xi_F(1) F(1)$, which equals unity in
the heavy quark limit.  The truncation errors in Table~5 are typically
less than half those of previous parameterizations, while the errors
in Table~6 are either better or worse by nearly a factor of two,
depending on the form factor\cite{BL} (we have corrected an oversight
in this reference, which used incorrect Blaschke factors for $F_0$ and
$G_0$).

	Note the especially small size of the truncation error for the
form factors $f_0$ and ${\cal F}_2$; if we consider all form factors
related by heavy quark symmetry, then the champion in this respect is
$S_{0+}$, with a truncation error of only 0.56\%.  However, as we
discuss in the next section, actually fitting data to such form
factors introduces much larger $1/m$ uncertainties.

\bigskip
\vbox{\medskip
\hfil\vbox{\offinterlineskip
\hrule
\halign{&\vrule#&\strut\quad\hfil$#$\quad\hfil\cr
height0pt&\omit&&\omit&&\omit&&\omit&&\omit&\cr
& F  && N_{\rm optimal} && \multispan3 \hfil {\rm Combined\ bounds\
from} $\bar B^{(*)} \bar D^{(*)}$ \hfil
  && T(\Xi_F F) & \cr
\noalign{\hrule}
& f_+ && 1.108  && -0.23 \le a_1 \le 0.20 &&
      -0.55  \le a_2 \le 0.58 && 2.6\% &\cr
& f_0 && 1.109  && -0.62 \le a_1 \le 0.58 &&
      -0.78 \le a_2 \le 0.85  && 0.7\% &\cr
\noalign{\hrule}
& f && 1.093 && -0.37 \le a_1 \le 0.39 &&
      -0.58 \le a_2 \le 0.56 && 1.8\% &\cr
& {\cal F}_1 && 1.093  && -0.06 \le a_1 \le 0.06 &&
      -0.11 \le a_2 \le 0.10 && 2.1\% &\cr
& g && 1.093 && -0.37 \le a_1 \le 0.40 &&
      -0.58 \le a_2 \le 0.57 && 1.2\% &\cr
& {\cal F}_2 && 1.093  && -0.41 \le a_1 \le 0.45 &&
      -0.59 \le a_2 \le 0.57 && 0.6\% &\cr
\noalign{\hrule}
& F_0 && 1.081 && -1 \le a_1 \le 1 &&
      -1    \le a_2 \le 1    && 6.3\% &\cr
& G_0 && 1.080 && -1 \le a_1 \le 1 &&
      -1    \le a_2 \le 1    && 9.4\% &\cr
} \hrule} \hfil
\medskip
\\
\INSERTCAP{5}{One-coefficient parameterizations using optimized
$N$. Bounds on $a_1$ ignore heavy quark violation. Bounds on $a_2$
allow for violation as described in Sec.\ \ref{add}. The truncation
error is relative to the normalization of $\, \Xi_F F$ at $w=1$.}}

	For baryon form factors, the large number of sub-threshold
poles typically ensures that at least two parameters are required.
Even with two parameters and the spin-symmetry improvements, the
truncation errors are significant. Using three parameters reduces the
truncation errors to negligible levels.

\bigskip
\vbox{\medskip
\hfil\vbox{\offinterlineskip
\hrule
\halign{&\vrule#&\strut\quad\hfil$#$\quad\hfil\cr
height0pt&\omit&&\omit&&\omit&&\omit&&\omit&\cr
& F  && N_{\rm optimal} && \multispan3 \hfil {\rm Combined\ bounds\
from} $\bar B^{(*)} \bar D^{(*)}$ \hfil
  && T(\Xi_F F) & \cr
\noalign{\hrule}
& F_1 && 1.104 && -0.90 \le a_1 \le 0.90 &&
     -0.97 \le a_2 \le 0.97 && 7.0\%  &\cr
& H_V && 1.104 && -0.31 \le a_1 \le 0.31 &&
     -0.53  \le a_2 \le 0.53 && 12\% &\cr
& G_1 && 1.104 && -0.98 \le a_1 \le 0.98 &&
     -0.99 \le a_2 \le 0.99 && 9.1\%  &\cr
& H_A && 1.104 && -0.15 \le a_1 \le 0.15 &&
     -0.26  \le a_2 \le 0.26 && 18\% &\cr
\noalign{\hrule}
& F_0 && 1.104 && -1 \le a_1 \le 1 &&
     -1 \le a_2 \le 1&& 0.35\%  &\cr
& G_0 && 1.104 && -1 \le a_1 \le 1 &&
     -1  \le a_2 \le 1 && 0.53\% &\cr
} \hrule} \hfil
\medskip
\\
\INSERTCAP{6}{Two-coefficient parameterizations using optimized
$N$. Bounds on $a_1$ ignore heavy quark violation. Bounds on $a_2$
allow for violation as described in Sec.\ \ref{add}. The truncation
error is relative to the normalization of $\, \Xi_F F$ at $w=1$.}}

     A way to circumvent the relatively large truncation errors on
most of the baryon form factors is revealed by an interesting feature
of Table~5: The $\Lambda_b \to \Lambda_c \ell \bar \nu$ form factor
$F_0$ is well-described (to about $6\%$) by a one-coefficient
parameterization. The contribution from $F_0$ to the decay rate is
suppressed by the lepton mass, so it is difficult to observe. However,
$F_0$ is related by heavy quark symmetry to all the other baryon form
factors [see (\ref{baryonhqs})]. Unlike in meson decays, the relation
including $1/m_c$ effects is known in terms of one constant\cite{GGW}
$\bar \Lambda_\Lambda \approx M_{\Lambda_b} - m_b$.  Thus, the
differential decay rate can be described in terms of two constants,
$a_1$ and $\bar \Lambda_\Lambda$. Using the value of $|V_{cb}|$
obtained from $\bar B \rightarrow D^{(*)} \ell \bar \nu$ and known
hadron masses should then allow a determination of $\bar \Lambda
\approx M_B - m_b$.  This quantity is important because it enters both
exclusive and inclusive semileptonic decay distributions as a $1/m_c$
correction.

	In addition to errors incurred by truncating the expansion
(\ref{xifexpansion}), there are a number of uncertainties arising from
various approximations we have made.  We enumerated in \cite{BGLNP}\ a
list of uncertainties which must be estimated for a reliable
determination of the quality of the dispersive bounds, and found that
we could allow for their effects by increasing truncation errors by
$40\%$.  Since then, we have greatly reduced or eliminated many of
these uncertainties.  In \cite{BL} we saw that the inclusion of the
additional parameter $N$ (or $t_0$) permits a dramatic reduction of
the truncation error, and the more realistic definition of the
truncation error used here reduces the uncertainties of one-parameter
fits even more.  The simultaneous inclusion of all $\bar B^{(*)} \bar
D^{(*)}$ states on the hadronic side of the dispersion relation serves
to help saturate the bound from the partonic side, and the explicit
inclusion of two-loop perturbative and leading nonperturbative effects
eliminates them as a source of uncertainty.  The uncertainty in $B_c$
pole positions is most significant for poles near threshold, and leads
to larger truncation errors only if the pole masses have been
overestimated. As pointed out in \cite{BL}, branch cuts from
multi-particle states below threshold can be ignored if they violate
isospin. The only uncertainty from \cite{BGLNP}\ that has not been
reduced is due to the choice of pole quark masses in the perturbative
calculation.  Combining the remaining uncertainties as in
\cite{BGLNP}, we find that their effects may be allowed for by
increasing $B \to D, D^*$ truncation errors by $20\%$ ({\it e.g.}, the
conservative truncation error for $f$ would be $2.2\%$) and $\Lambda_b
\to \Lambda_c$ truncation errors by $30\%$. In nearly all cases, this
small increase makes no practical difference.

\section{Slope and Curvature Relations} \label{CNsec}

	A set of interesting relations between the slopes and
curvatures of \btod\ and \btods\ form factors has been derived by
Caprini and Neubert\cite{CN}. Here we examine these relations in the
context of the parameterization formalism, point out and circumvent an
invalidating assumption, and discuss the utility of the new, valid,
relations.

	We have seen that each of the form factors $F$ for $\bar
B^{(*)} \rightarrow D^{(*)} \ell \bar \nu$ can be expanded in a series
\begin{equation}
F(z) = {1 \over P(z) \phi(z) } \sum_{n=0}^\infty a_n \, z^n \ \ ,
\end{equation}
where, in order to compare with~\cite{CN}, we have set $t_0 = t_-$
($N=1$). From Eq.~(\ref{abound}), the coefficients $a_n$ obey
\begin{equation}\label{cnabound}
a_0^2 + a_1^2 + a_2^2 \le 1  .
\end{equation}
Expressing $a_0, a_1$, and $a_2$ in terms of $z$-derivatives of $P(z)
\phi(z) F(z)$ at $z=0$, Eq.~(\ref{cnabound}) gives
\begin{equation} \label{ellipseq}
\left. \left[ P(z) \phi(z) F(z) \right]^2 \right|_{z=0} + 
\left. \left[ {d\over dz}\left( P(z) \phi(z) F(z) \right) \right]^2
\right|_{z=0} +
\left. \frac 1 4 \left[ {d^2\over dz^2} \left( P(z) \phi(z) F(z)
\right) \right]^2 \right|_{z=0} \le 1 \\ .
\end{equation}
For $t_0=t_-$, $z=0$ corresponds to $w=1$. Then Eq.~(\ref{ellipseq})
constrains the slope $-\rho^2$ and curvature $c$ defined by
\begin{equation}
 F(w) = F(1) - \rho^2 (w-1) + c \, (w-1)^2 + O[(w-1)^3] ,
\end{equation}
to lie within an ellipse.

	Equation (\ref{ellipseq}) is the starting point of
reference\cite{CN}, with one critical simplification: They consider
the form factor $\tilde f_0 \equiv f_0/[(M_B - M_D) \sqrt{M_B M_D}
(1+w)]$, and argue that it does not receive contributions from scalar
$B_c$ mesons that would generate poles in $f_0$.  This allows them to
set $P(z) =1$, leading to a nearly linear relation between $\rho^2$
and $c$,
\begin{equation} \label{crhowrong}
c \approx  0.72 \rho^2 - 0.09 .
\end{equation}

	Their argument is based on the assumption that the scalar
$B_c$ mesons are broad resonances because they can decay into
two-particle intermediate states such as $B_c (0^-) + \eta$.
Unfortunately, the very potential models they cite refute this
idea. For example, Ref.~\cite{Quigg} has the mass of the scalar $2\ {
}^3 P_0$ state, 6.700 GeV, below the two-particle threshold for an
$\eta$ plus the $1\ { }^1 S_0$ $B_c$ ground state, 6.264 + 0.548 =
6.812 GeV. The scalar $2\ { }^3 P_0$ is thus essentially stable with
respect to hadronic transitions, since transitions involving one pion
are suppressed by isospin, two pions by parity (or phase space, in
decays to a $B_c^*$), and three pions by phase space. All the
references\cite{pot} that calculate the relevant scalar masses and
widths agree that there are two narrow, scalar $B_c$ resonances below
the $\bar B \bar D$ threshold.  It is also worth nothing that the
lowest-lying charmonium scalar state $\chi_{c0}$ is narrow.

	While sub-threshold branch cuts from states containing at
least a $b$ and a $c$ quark may be legitimately
ignored\cite{BL,BGLNP}, it is well known that poles play an essential
role in the shape of the form factor\cite{BS,spoilers}.  For this
reason, the slope-convexity relations derived in~\cite{CN} are
invalid.

	New relations can be derived by simply including the two
scalar $B_c$ states in $P(t)$. We use masses from\cite{Quigg}, which
agree with other potential model determinations to better than 1\%.
It is algebraically straightforward to input physical masses and
expand our parameterization in powers of $(w-1)$,
\begin{eqnarray} \label{fzeroexp}
\tilde f_0(w) & = &  \tilde f_0(1) \cr 
&+& \left(+1.72 a_1 - 0.77 \tilde f_0(1) \right) (w-1) \cr
&+&\left(-1.74 a_1 + 0.21 a_2 + 0.55 \tilde f_0(1) \right) (w-1)^2 \cr
&+& \left(+1.41 a_1 - 0.27 a_2 -0.38 \tilde f_0(1) \right) (w-1)^3 \cr
&+& \left(-1.03 a_1 + 0.25 a_2 + 0.25 \tilde f_0(1) \right) (w-1)^4 +
\ldots ,
\end{eqnarray}
and solve for the coefficient $c$ of $(w-1)^2$ in terms of the
coefficient $ - \rho^2$ of $(w-1)$. We find
\begin{equation}\label{crho}
 c = 1.02 \rho^2 + 0.21 a_2 - 0.23 \tilde f_0(1) .
\end{equation}
This is not a very interesting relation because the unknown
coefficient $a_2$, which can be as large as $\pm 1$ ($\approx \pm 0.6$
if we include the contribution of higher states and ignore heavy quark
symmetry violation), significantly affects the slope-convexity
relationship.  Had we ignored the Blaschke factor $P(t)$, the
coefficient of $a_2$ would have been $0.07$.

	For $N=1$, our usual truncation analysis shows that $a_2$ can
contribute at most $4\%$ to $\tilde f_0(w)$. This may seem surprising,
given its obvious importance in (\ref{fzeroexp}) and (\ref{crho}).
The $4\%$ value arises from a cancellation of the $a_2$ dependence
among the various $(w-1)^n$ coefficients; note the alternating signs
of the $a_n$ coefficients.  The cancellation is not accidental, but
reflects the naturalness of an expansion in $z(w;N)$ rather than
$(w-1)$.  This effect is highlighted by the observation that the
$(w-1)^3$ term in (\ref{fzeroexp}) can be as large as $40\%$ at
$w_{\rm max}=1.6$, indicating that $\tilde f_0$ must be expanded to
rather high order in $(w-1)$ if percent-level accuracy is desired.

	The expansion (\ref{fzeroexp}) can, alternatively, be used to
test heavy quark symmetry by placing a restriction on the slope
$-\rho^2$. If we include the contribution from the spin-related states
$\bar B^{(*)} \bar D^{(*)}$, $a_1$ is restricted to $-0.61 \le a_1 \le
0.59$, leading to
\begin{equation}
-0.26 \le \rho^2 \le 1.84 \ \ .
\end{equation}
The same relation can be derived using the form factors $S_{0+}$ or
$S_{00}$.  These bounds are somewhat weaker than those derived from
Bjorken\cite{BJ} and Voloshin\cite{Volo} inequalities\footnote{The
physics leading to these results is quite different: The Bjorken and
Voloshin inequalities use perturbative QCD to bound exclusive form
factors directly in the semileptonic region, while the dispersion
constraints use both perturbative QCD in the pair-production region
and the phenomenological mass spectrum in the unphysical region $t_- <
t < t_+$.}, which restrict $ 0.22 \le \rho^2 \le 1.15$ once
$O(\alpha_s)$ corrections have been included\cite{alphaBJ}.

\section{Experimental Fits} \label{fit}

	While constraints obtained from unobserved form factors like
$f_0$ may serve as tests of heavy quark symmetry, they are not
well-suited for fitting to data. The reason is that once the
truncation error on a form factor is sufficiently small (a few
percent), heavy quark symmetry violating effects of order 20--30\%
become the main concern. For \btod, using the parameterization of
$f_+$ avoids any dependence on heavy quark symmetry.  For \btods,
using the form factor $f$ minimizes the dependence on heavy quark
symmetry. This is because on the one hand, the ratio
\begin{equation}
{f \over M_B \sqrt{M_B M_{D^*} } (1+w) g} = 
   {\hat C^5_1 \over  \hat C_1} \Bigl[ 1 
   + {\bar \Lambda \over 2 m_c} (w-2) \Bigr] 
      +  O \left( \frac1{m_b}, \frac1{m_c^2} \right) .
\end{equation}
is given in terms of a single\cite{Luke}, roughly determined constant
$\bar \Lambda \approx 300$--$600\ {\rm MeV}$ and known\cite{benNeub}
perturbative functions $ C^5_1/C_1 = 1 + O(\alpha_s)$, while on the
other hand the ratio $a_+/g= -1/2$ is determined using only spin
symmetry, which is expected to hold more precisely than full
flavor-spin symmetry\cite{bgrin}. We use $f$ rather than $g$ because
it is protected from $1/m$ corrections\cite{Luke} at zero recoil.

	The purpose of the following fits is not to extract the best
value of $|V_{cb}|$, since only the experimental groups themselves can
correctly account for efficiencies, resolutions, smearing effects,
etc.  Rather, since we expect the approximate results from using the
QCD-derived parameterization to survive these experimental
corrections, these fits may be used to motivate a more thorough
analysis.

      Form factors for $\Lambda_b \rightarrow \Lambda_c \ell \bar \nu$
may be extracted in the near future at CDF\cite{cdf} or
LEP\cite{alephlam}.  Current data for $\bar B \rightarrow D^{(*)} \ell
\bar \nu$ decay spectra are available from CLEO\cite{cleo},
ALEPH\cite{aleph}, OPAL\cite{opal}, and DELPHI\cite{delphi}; older
data exists from ARGUS\cite{argus}.  Some progress has been made
towards measuring individual form factors in
\btods\cite{cleo2}. When this is finally accomplished, the
parameterizations for individual form factors can be applied without
recourse to heavy quark symmetry, except for the uncertainties in the
value of $F(1)$.

	In the meantime, one must rely on the heavy quark symmetry
prediction that the \btods\ differential rate is proportional to a
function ${\cal F} (w)$ that is normalized to $1 + O(1/m^2)$ at zero
recoil and is proportional to the Isgur-Wise function $\xi(x) =
\Xi_f(w) f(w)$ up to $1/m_c$ corrections. We may then use the
one-coefficient, QCD-derived parameterization of $f$ obtained from
(\ref{mesonhqs}) and (\ref{xifexpansion}) to extract from data the
values of $|V_{cb}| {\cal F} (1)$ and $a_1/{\cal F} (1)$,
\begin{eqnarray}
{ {\cal F} (w) \over {\cal F} (1) } &=&
  {1 \over (1+w) P_f(w) \phi_f(w;N)  } \left\{
  2 P_f (1) \phi_f (1;N) + {a_1 \over {\cal F} (1)}
  {\left[z(w;N) - z(1;N) \right] \over \sqrt{M_B M_{D^*}}} \right\},
\end{eqnarray}
where $N= 1.093$, $z(w;N)$ is defined in Eq.~(\ref{zofw}),
$\phi_f(w;N)$ is given by Eq.~(\ref{phiofz}) and Table~1, and $P_f(w)$
is determined from Eqs.~(\ref{Poft}), (\ref{wdef}), and the first four
vector masses in Table~3.

	The procedure is precisely as detailed in \cite{BGLNP}, except
that we now have only one fit coefficient instead of two\cite{BL}.
Morever, the improvements described above reduce our one-parameter
truncation errors to no more than 3\%.  We fit to the experiments
whose differential distributions are easily available.  A $\chi^2$ per
degree of freedom (d.o.f.) fit using our QCD dispersion relation
bounds (QCD fit) to CLEO data\cite{cleo}\ gives 
\begin{eqnarray} 
{\rm QCD\ Fit} \qquad 
   10^3 \cdot |V_{cb}| \ {\cal F} (1) &=&  36.9^{+2.0}_{-2.1} \ ,
\nonumber \\
   \frac{a_1}{ {\cal F} (1) } &=& 0.000^{+0.022}_{-0.019} \ , \\
{\rm Linear\ Fit} \qquad 
   10^3 \cdot |V_{cb}|\ {\cal F} (1) &=&  35.1^{+1.9}_{-1.9} \ , 
\end{eqnarray}
with $\chi^2_{\rm min}/{\rm d.o.f.} = 0.67$, while a fit to ALEPH
data\cite{aleph}\ gives
\begin{eqnarray}
{\rm QCD\ Fit} \qquad 
    10^3 \cdot |V_{cb}| \ {\cal F} (1) &=&  31.9^{+2.4}_{-2.4} \ ,
\nonumber \\
     \frac{a_1}{ {\cal F} (1) } &=& 0.096^{+0.040}_{-0.034} \ , \\ 
{\rm Linear\ Fit} \qquad
    10^3 \cdot |V_{cb}| \ {\cal F} (1) &=&  31.9^{+1.8}_{-1.8} \ ,
\end{eqnarray}
with $\chi^2_{\rm min}/{\rm d.o.f.} = 0.74$.  We have also included
the values quoted by the experimental groups using a linear fit for
comparison, and only statistical errors at one standard deviation
are listed.

	For \btod, we can fit the parameterization of $f_+$,
\begin{eqnarray}
{ f_+(w) \over f_+(1)} &=&
  {1 \over P_{f_+}(w) \phi_{f_+}(w;N)} \left\{
  P_{f_+}(1) \phi_{f_+}(1;N) + { a_1 \over f_+(1)}
  \left[z(w;N) - z(1;N) \right] \right\} ,
\end{eqnarray}
directly to data, without the need to invoke heavy quark symmetry.
Here $N= 1.108$ and $P_{f_+}$ depends on the first three vector
masses in Table~3.

	In this case, a fit to ALEPH data\cite{aleph}\ yields 
\begin{eqnarray}
{\rm QCD\ Fit\ } \qquad
    10^3 \cdot |V_{cb}| {\cal F}_D (1)   
     &=&  29.2^{+7.3}_{-7.3} \ , \nonumber \\
     {a_1 \over {\cal F}_D (1)} &=& 0.095^{+0.110}_{-0.066} \ ,\\
{\rm Linear\ Fit\ } \qquad
    10^3 \cdot |V_{cb}| \ {\cal F}_D (1) &=&  27.8^{+6.8}_{-6.8} \ , 
\end{eqnarray}
with\footnote{The larger $\chi^2/{\rm d.o.f.}$ is due to their binned
data point at $w = 1.55$, which suggests a peculiar upturn of the form
factor near $w_{\rm max}$.}  $\chi^2_{\rm min}/{\rm d.o.f.} = 1.94$,
and now ${\cal F}_D(1) = \Xi_{f_+} (1) f_+ (1) = 2 \sqrt{r}
f_+(1)/(1+r)$.

	For CLEO \btods\ data, which exhibits discernible curvature,
our central values of $|V_{cb}| {\cal F} (1)$ lie at the upper
1$\sigma$ boundary of the linear-fit result, while for ALEPH data,
which is extremely flat, the central values are nearly the same. 
The statistical errors are larger because the QCD parameterization
allows for curvature.

      The ALEPH \btod\ data presents an interesting area in which to
test our results: The shape of their data will have to change as the
statistical errors are reduced, if it is to be consistent with QCD.

\section{Conclusions} \label{conc}

	Form factors can be reliably bounded in the pair-production
region of momentum-space by perturbative QCD calculations.
Analyticity, crossing symmetry, and dispersion relations may then be
used to translate these bounds into constraints in the
phenomenologically interesting semileptonic region.

	While these constraints typically imply rather weak bounds on
the slopes of form factors\cite{BGLPRL,CM,CM2,Taron}, quite stringent
bounds can be obtained if the form factor at two or more points is
known. The constraints actually imply an infinite number of
increasingly stringent bounds, depending on the number of points at
which the form factor is known\cite{BGLPRL,Lel}. All of these bounds
are automatically obeyed if the form factor is parameterized as in
Eq.~(\ref{masterparam}), even if the parameterization is truncated
after a few terms.

	In this paper, we have eliminated some of the uncertainty
involved in the derivation of these parameterizations by including
two-loop perturbative corrections to the partonic side of the
dispersion relation. We have also presented parameterizations for form
factors whose contribution to semileptonic decay rates is suppressed
by the lepton mass.

	We examined a relation between the slope and curvature of the
Isgur-Wise function derived by Caprini and Neubert\cite{CN}, and
pointed out the presence of sub-threshold singularities that
invalidate their analysis.  Once these singularities are correctly
accounted for, the slope-convexity relations become rather weak.
Bounds on the slope of the Isgur-Wise function made by ignoring finite
mass corrections are also fairly weak.  We point out that even with
strong constraints on the slope and curvature, higher-order terms in a
$(w-1)$ expansion of the Isgur-Wise function can be quite large. The
parameterization in Eq.~(\ref{masterparam}) does not suffer from this
limitation.

	For $b \to c$ transitions, we reduced the {\it truncation
errors} that describe the accuracy of such parameterizations.  This
was accomplished in part by using a parameterization that is
automatically normalized at zero recoil to a quantity ${\cal F}(1)$,
which must be supplied by some other method such as heavy quark
symmetry. More importantly, we included the contributions of higher
states in the hadronic side of the dispersion relation. These states,
the $\bar B^* \bar D$ and $\bar B^*\bar D^*$ pairs, are related to
\btod\ and \btods\ form factors in the semileptonic region by heavy
quark symmetry.  Even in the presence of substantial heavy quark
symmetry violation, these relations place lower bounds on the
contribution from the higher states to the dispersion relation that
lead to tighter upper bounds on the magnitudes of the unknown
parameterization coefficients.  This in turn reduces the truncation
errors on the parameterizations of the various form factors.

	For most of the $\bar B \to D, D^*$ form factors, the
inclusion of higher states reduces truncation errors by roughly
$40\%$.  In one case, ${\cal F}_1$, the truncation error is reduced by
as much as a factor of ten. This hefty improvement arises because
${\cal F}_1$ contributes very little to the dispersion relation, so it
is far from saturating the perturbative bound until its spin-symmetry
partners are included as well. After including all $\bar B^{(*)} \bar
D^{(*)}$ states, we find that each of the six form factors governing
\btod\ and \btods\ is described to better than $3\%$ accuracy using
only one unknown parameter. This should be a considerable aid in
experimentally disentangling the various form factors in differential
decay distributions.

	For $\Lambda_b \to \Lambda_c \ell \bar\nu$ decays, there
are no spin-symmetry partners to help saturate the dispersion relation
bound. However, the presence of more than one helicity amplitude in
the same bound achieves the same effect. This is most dramatic in the
case of $H_A$, whose truncation error is reduced by a factor of four.
Of greater interest is the observation that the form factor $F_0$ can
be described at the $6\%$ level using only one unknown coefficient.
Because the $1/m_c$ corrections to heavy quark symmetry relations
among the baryonic form factors are given in terms of one additional
parameter, $\bar \Lambda_\Lambda$, the entire decay distribution can
be described using only two unknown constants. This should allow a
relatively clean extraction of the phenomenologically interesting
quantity $\bar \Lambda_\Lambda$.

	Finally, we used a parameterization of the form factor $f$ and
heavy quark relations to extract $|V_{cb}|$ from \btods\ data. This
choice of form factor minimizes the dependence on heavy quark
symmetry. We expect the qualitative features of this extraction to
persist even after the effects of experimental resolution, smearing, 
etc., are properly incorporated. Our analysis suggests that the
implicit error associated with the choice of parameterization 
is comparable to the statistical errors normally quoted.  
Similar statements apply to \btod,
except that in this case no reliance on heavy quark symmetry is
necessary, since we can parameterize the form factor $f_+$ directly.

	Further improvements may be possible by including additional
higher states in the dispersion bound, or perhaps by weighing the
dispersion integral differently. One could also readily incorporate
approximate $SU(3)$ symmetry by using an effective $n_I = 2.5$ isospin
factor\cite{CN} in $\bar B \to D, D^*$ decays, which would decrease
truncation errors by an additional $10\%$.  While such improvements
would be welcome for the baryonic form factors, their utility for
$\bar B \to D, D^*$ form factors is not clear. This is because, once
truncation errors are at the few-percent level, the overwhelming
source of uncertainly comes from heavy quark symmetry violations,
which are expected to be of order $30\%$.  Such uncertainties
highlight the importance of extracting individual form factors, which
can be parameterized using one coefficient without recourse to heavy
quark symmetry.

\vskip1.2cm
{\it Note Added}
\hfil\break
The essential role of the scalar $B_c$ poles in the slope-convexity
relations of Caprini and Neubert has been pointed out independently by
L. Lellouch ({\it private communication}). A corrected slope-convexity
relation and other topics related to those in this work are in
preparation by these authors\cite{CLN}.  Additional criticism of the
neglect of the scalar poles appears in Refs.~\cite{Ural}.

\vskip1.2cm
{\it Acknowledgments}
\hfil\break
We would like to thank Andr\'{e} Hoang, Bernd Kniehl, Karl Schilcher,
and Sotos Generalis for discussions concerning higher-order QCD
results, Aneesh Manohar for discussions about the partial wave
decomposition, and Martin Savage and Lawrence Gibbons for 
comments on statistical errors.  
This work is supported by the Department of Energy
under contracts Nos.\ DOE-FG03-97ER40506 and DOE/ER/40682-132.

\end{document}